\documentclass[12pt]{article}

\usepackage{amsfonts,amssymb,amsmath,exscale,relsize,hyperref,enumitem,amsthm,mathtools, physics}
\usepackage{graphicx}
\usepackage{float}
\usepackage[thinlines]{easytable}
\usepackage{csquotes}

\usepackage[authoryear]{natbib} 
\usepackage{hyperref}           
\usepackage{varioref}
\usepackage{subcaption}
\usepackage{xcolor}
\usepackage{setspace}
\usepackage{appendix}
\usepackage[margin=1.35in]{geometry}
\usepackage{setspace} 

\newtheorem{theorem}{Theorem}

\newtheorem{lemma}{Lemma}

\newtheorem{proposition}{Proposition}

\def\T{\mathrm{\scriptscriptstyle{T}}}
\def\HB{\mathrm{\scriptscriptstyle{HB}}}
\def\PHB{\mathrm{\scriptscriptstyle{PHB}}}
\def\H{\mathrm{\scriptscriptstyle{H}}}
\def\PH{\mathrm{\scriptscriptstyle{PH}}}

\mathchardef\mhyphen="2D

\def\T{{\mathrm{\scriptscriptstyle T}}}
\def\HB{{\mathrm{\scriptscriptstyle HB}}}
\def\PHB{{\mathrm{\scriptscriptstyle PHB}}}
\def\ML{{\mathrm{\scriptscriptstyle ML}}}
\def\JS{{\mathrm{\scriptscriptstyle JS}}}

\def\T{{ \mathrm{\scriptscriptstyle T} }}

\title{On Quantification of Borrowing of Information in Hierarchical Bayesian Models}

\author{%
	Prasenjit Ghosh\thanks{Department of Statistics, Texas A\&M University, College Station, TX 77843, USA. Email: \texttt{prasenjit@stat.tamu.edu}}%
	\hspace{1em} 
	Anirban Bhattacharya\thanks{Department of Statistics, Texas A\&M University, College Station, TX 77843, USA. Email: \texttt{anirbanb@stat.tamu.edu}}%
	\hspace{1em} 
	Debdeep Pati\thanks{Department of Statistics, University of Wisconsin--Madison, Madison, WI 53706, USA. Email: \texttt{dpati2@wisc.edu}}%
}

\date{} 

\newcommand{\keywords}[1]{%
	\par\addvspace\baselineskip
	\noindent\textbf{Keywords: }#1
}

\begin{document}
	
	\maketitle
	
	\begin{abstract}
		In this work, we offer a thorough analytical investigation into the role of shared hyperparameters in a hierarchical Bayesian model, examining their impact on information borrowing and posterior inference. Our approach is rooted in a non-asymptotic framework, where observations are drawn from a mixed-effects model, and a Gaussian distribution is assumed for the true effect generator. We consider a nested hierarchical prior distribution model to capture these effects and use the posterior means for Bayesian estimation. To quantify the effect of information borrowing, we propose an integrated risk measure relative to the true data-generating distribution. Our analysis reveals that the Bayes estimator for the model with a deeper hierarchy performs better, provided that the unknown random effects are correlated through a compound symmetric structure. Our work also identifies necessary and sufficient conditions for this model to outperform the one nested within it. We further obtain sufficient conditions when the correlation is perturbed. Our study suggests that the model with a deeper hierarchy tends to outperform the nested model unless the true data-generating distribution favors sufficiently independent groups. These findings have significant implications for Bayesian modeling, and we believe they will be of interest to researchers across a wide range of fields.
	\end{abstract}
	
	
	\section{Introduction}
	Hierarchical Bayes models have long been recognized as effective and powerful modeling tools for combining information across multiple sources, such as in multi-center studies and meta-analyses. They offer great flexibility to accommodate prior information by breaking it down into several layers of conditional distributions. Such models are particularly valuable for modeling large and complex datasets found in genome-wide association studies, proteomics, clinical trials, bio-informatics, covariance matrix estimation, image retrieval, longitudinal studies, and unsupervised, semi-supervised, or reinforced learning problems. Early accounts of hierarchical Bayes models can be found in various sources, including \cite{Good_1950, Good_etal_1966, Good_1980} and \cite{Lindley_Smith_1972}. For more recent and in-depth treatments, please refer to \cite{Berger_2013, Bernardo_Smith_2009, Chen_etal_2011, Dunson_2010, Gelman_etal_2014, Griffin_Holmes_2010, Lehmann_Casella_2006, Robert_2007}, only to name a few.  
	
	A central feature of hierarchical Bayes models is their ability to borrow strength by sharing information through shared hyperparameters or latent factors. Originating from ideas first proposed by John W. Tukey, this principle has become fundamental not only in traditional statistical modeling but also in multi-task learning, meta-learning, and transfer learning \citep{Allen_etal_2018, Amit_Meir_2018, Finn_Abbeel_Levine_2017, Grant_etal_2018, Gordon_etal_2019, Hu_etal_2019, Karbalayghareh_etal_2018, Kong_etal_2020,  Ravi_Beatson_2019, Yoon_etal_2018}. It is particularly useful when dealing with a large number of parameters, and a limited training sample is available to estimate each parameter. In meta-learning, for instance, where the focus is to facilitate rapid learning of a novel future task after extracting knowledge about several other similar tasks, the data is presented in a two-level hierarchy, allowing the learner to accumulate knowledge in a way that captures the common structure across tasks while remaining flexible enough to adapt to new ones. A similar mechanism applies to multi-task learning, where the shared statistical structure is leveraged to solve multiple observed tasks. These hierarchical strategies have been proven effective in both supervised and reinforcement learning, leading to prominent approaches such as few-shot learning, PAC Bayesian Analysis, and hierarchical Bayes Model-Agnostic Meta-Learning (MAML).

	The purpose of this paper is to explore the effects of sharing information across multiple layers of learning and to analyze the role of latent hierarchies or shared hyperparameters on such information sharing. Previous literature has shown that the gain in information about the hyperparameters decreases as one moves to the bottom of the hierarchy \cite{Goel_1983, Goel_Degroot_1979, Goel_Degroot_1981}. Non-parametric Bayes methods have been explored for hierarchical borrowing of information across different studies, centers, or exchangeable groups of data \cite{Dunson_2010}. In a recent study \cite{Nguyen_2016}, posterior contraction rates of the base probability measure of a hierarchical Dirichlet process before inference in multiple groups of data \citep{Teh_etal_2005} have been found. Bayesian hierarchical models have also been proposed for subgroup analysis \cite{Xu_etal_2020}, with two metrics based on Mallow's distance \cite{Mallows_1972} used to determine the borrowing of strengths across different subgroups and the overall borrowing of strength. In this work, we consider a non-asymptotic framework where observations are sampled from a mixed-effects model, and the true effect-generating distribution is assumed to be Gaussian, characterized by a specific correlation structure. In the Bayesian literature, such mixed-effects models are quite popular for analyzing longitudinal data \citep{Chung_Cho_2022, Daniels_Hogan_2008, Diggle_2002, Fitzmaurice_Laird_Ware_2011, Hedeker_Gibbons_2006, Laird_Ware_1982, Molenberghs_Verbeke_2005, Schiratti_etal_2017, Verbeke_Molenberghs_2009, Verbeke_etal_2014}, multilevel or hierarchical data \citep{Gelman_Hill_2007, Goldstein_2011, Hox_etal_2017, Ohlssen_2007, Snijders_Bosker_2012} and repeated measure designs \citep{Laird_Ware_1982, Lindsey_1993, Molenberghs_etal_2010, Raudenbush_Bryk_2002, Salakhutdinov_etal_2013}. These statistical models incorporate both fixed and random effects, which provide them with a hierarchical structure. Fixed effects describe the data at the population (or group) level, while random effects describe the data at the individual level. We consider a pair of nested hierarchical Bayes priors to model these effects and refer to the model nested inside the one with a deeper hierarchy as the \textit{partial--hierarchical Bayes} model. In the Bayesian paradigm, parameters are treated as fixed but unknown points, and the likelihood-prior combination yields a posterior distribution, which serves as a measure-valued estimate. The efficacy of this approach is quantified through classical risk measures. Existing literature shows that Bayes estimators obtained under the aforesaid hierarchical Bayes models cannot be regarded as a uniform improvement over the other in terms of the classical risk measures \citep{Lehmann_Casella_2006}. Moreover, intricate analytic expressions of the classical risk functions of these estimators \citep{ Bock_1975, Egerton_Laycock_1982} render rigorous theoretical comparisons between their classical risk profiles. This limitation complicates the process of drawing clear theoretical conclusions regarding the advantages of additional hierarchy.
	
	In contrast, our approach in this article encompasses an integrated measure of risk obtained by averaging the classical squared $\ell_2$ risk function with respect to the generative distribution of the random effects. We compare the integrated risk profiles of these Bayes estimators and propose to quantify the effect of borrowing information through the gain in risk for the hierarchical Bayes estimator compared to its partial-hierarchical Bayes counterpart. Thus, the methodology utilized in the present analysis is unique as it involves the assumption of a generative distribution for the parameter and the subsequent integration over it to define an integrated measure of risk, enabling differentiation between procedures. Therefore, our current approach offers an alternative perspective on the analysis of mixed-effects models, enabling a more nuanced assessment of the efficacy of various procedures. By averaging the risk with respect to a generative distribution of random effects, the integrated risk provides a coherent decision-theoretic criterion. This integrated measure captures the overall quality of estimation in a non-asymptotic sense and offers a principled approach to quantifying the benefits of information borrowing across different levels of hierarchy—an insight that classical risk analysis alone cannot provide.
	
	
	
	Our focus in this article is on determining precise analytical conditions for the true data-generating mechanism under which the hierarchical Bayes estimator can outperform its competitors. We consider two scenarios where the unknown effects are correlated: (i) a compound symmetric structure and (ii) a perturbation of the compound symmetry. In scenario (i), we establish the necessary and sufficient conditions under which the hierarchical Bayes estimator is superior. A similar set of sufficient conditions is also derived for scenario (ii). These conditions ensure that a model with a deeper hierarchy will outperform a nested model, unless the true data-generating distribution indicates that the random effects are very weakly correlated.

	Our results indicate that, except in rare cases where the true random effects are nearly independent, the hierarchical Bayes estimator can achieve a lower integrated risk compared to its partial-hierarchical counterpart. This finding provides a formal justification for the common belief that deeper hierarchical structures promote more effective information sharing across groups. In practical terms, the analysis suggests that practitioners modeling correlated populations—such as in multi-center clinical trials or meta-analyses—should prefer richer hierarchical specifications. The benefits of borrowing information in these cases may outweigh the minor risk of misspecification that comes with adding another level of hierarchy.

	The remainder of the paper is structured as follows. Section 2 introduces the hierarchical Bayesian models for the normal means setting and defines the integrated risk. Section 3 develops the main theoretical results, followed in Section 4 by simulation studies that validate and illustrate these findings. Concluding remarks and potential directions for future research are provided in Section 5. Finally, the Appendix contains the proofs of the theoretical results along with several auxiliary results.
	
	
	
	\section{Hierarchical Bayes in Normal-means Model}
	Consider the problem of joint estimation of $d$ unknown parameters $\{\mu_j\}_{j=1}^d$ in a normal-means model  \citep{Johnstone_Silverman_2004}:
	\begin{align}
		Y_{ij} \mid \mu_{j} \stackrel{}{\sim} N(\mu_{j},1),\ \textrm{independently for } i=1,\ldots,n,\textrm{ and } j=1,\ldots,d. \label{TRUE_MODEL}
	\end{align}
	Henceforth we use the symbol $Y$ to denote the complete data vector $(Y_{11},\dots,Y_{n1},\dots,Y_{1d}$, $\dots,Y_{nd})$. There is an enormous literature associated with the estimation of $\mu = (\mu_1, \ldots, \mu_d)^\T$ in the above normal--means model \eqref{TRUE_MODEL}. See, for instance, \citep{Efron_2012, Efron_Morris_1971, James_Stein_1961, Johnstone_Silverman_2004, Lehmann_Casella_2006, Stein_1981}, and references therein. The classical James-Stein phenomenon exhibits inadmissibility of the maximum likelihood estimator for $d \ge 3$ under the quadratic error loss
	\small\begin{equation}\label{LOSS_FUNCTION}
		\|\hat{\mu}-\mu \|^2_{2}=\sum_{j=1}^{d}(\hat{\mu}_{j}-\mu_{j})^2.
	\end{equation}\normalsize
	Specifically, let $R(\hat{\mu}, \mu) = E_{Y\mid \mu} \|\hat{\mu} - \mu\|_2^2$ denote the squared $\ell_2$ risk of an estimator $\hat{\mu}$ assuming the true parameter vector to be $\mu$, where $E_{Y \mid \mu}$ represents the expectation concerning the true data generating distribution. Then, it is well known that $R(\hat{\mu}_{\JS}, \mu) \le R(\hat{\mu}_{\ML}, \mu)$ for all $\mu \in \mathbb{R}^d$, where $\hat{\mu}_{\JS}$ and $\hat{\mu}_{\ML}$ respectively denote the James--Stein and the maximum likelihood estimators of $\mu$. The James--Stein estimator has a natural empirical Bayesian interpretation \citep{Efron_Morris_1972} under conditionally independent Gaussian priors
	\begin{equation}\label{PHB_MODEL}
		\mu_{j} \mid \tau^2  \stackrel{}{\sim} N(0,\tau^2), \ \textrm{independently for } j=1,\ldots,d.
	\end{equation}
	We refer to (\ref{PHB_MODEL}) as a partial-hierarchical Bayes (PHB) model. Using the conjugacy properties of Gaussian priors, standard statistical calculations yield the following James--Stein type empirical Bayes estimator of each $\mu_{j}$ \citep{Efron_2012, Lehmann_Casella_2006} under model (\ref{PHB_MODEL}) as\small
	\begin{equation}\label{PHB_ESTIMATES}
		\hat{\mu}_{j,\PHB}=\bigg[1-\frac{d-2}{n\sum_{j=1}^{d}\bar{Y}_{j}^2}\bigg]\bar{Y_{j}},
	\end{equation}\normalsize
	$\mbox{provided } d \geq 3.$ In (\ref{PHB_ESTIMATES}) above, $\bar{Y}_{j}=n^{-1}\sum_{i=1}^{n}Y_{ij}$ denotes the sample mean for the $j-$th normal population. 
	
	Our goal here is to compare two different hierarchical Bayesian models concerning the estimation of $\mu$. Towards that, the unknown $\mu_{j}$s are modeled through a hierarchical Bayes (HB) prior distribution as
	\begin{equation}\label{HB_MODEL}
		\begin{aligned}
			\mu_{j}\mid\eta,\tau^2  &\stackrel{}{\sim} N(\eta,\tau^2), \ \textrm{independently for } j=1,\dots,d,\\
			\eta  &\sim \mbox{ } \pi(\eta)\equiv 1, \eta\in\mathbb{R}.
		\end{aligned}
	\end{equation}
	The corresponding empirical Bayes estimators of $\mu_{j}$ \citep{Efron_2012, Lehmann_Casella_2006} under model (\ref{HB_MODEL}) is given by\small
	\begin{equation}\label{HB_ESTIMATES}
		\hat{\mu}_{j,\HB}=\bar{Y}+\bigg[1-\frac{d-3}{n\sum_{j=1}^{d}(\bar{Y}_{j}-\bar{Y})^2} \bigg](\bar{Y}_{j}-\bar{Y}),
	\end{equation}\normalsize
	$\mbox{provided } d \geq 4$, where $\bar{Y}=d^{-1}\sum_{i=1}^{d}\bar{Y}_{j}$ denotes the grand mean of the $Y$'s.
	
	In model (\ref{HB_MODEL}), we have a shared location hyperparameter, namely, $\eta$, among the independent prior specifications of the $\mu_{j}$'s, which is further assigned a uniform non-informative hyperprior $\pi(\eta)$ at the next level of hierarchy. In both models (\ref{PHB_MODEL}) and (\ref{HB_MODEL}), the variance component $\tau^{2}$ is assumed to be unknown \textit{a priori} but is not assigned any additional hyperprior. However, for inference, $\tau$ is replaced by its standard empirical Bayes estimators depending on what prior specification between (\ref{PHB_MODEL}) and (\ref{HB_MODEL}) we choose. Here model (\ref{PHB_MODEL}) is nested within model (\ref{HB_MODEL}) (for instance, take $\pi(\eta)$ to be degenerate at $0$ in model (\ref{HB_MODEL})).\par
	
	Our aim here is to investigate theoretically the effect of information borrowing through $\eta$ on the joint estimation of $\mu_{j}$s by considering the partially hierarchical Bayes model (\ref{PHB_MODEL}) as a \textit{baseline model} and comparing the integrated risk profiles of $\hat{\mu}_{\HB}$ and $\hat{\mu}_{\PHB}$. When $\mu$ is assumed to be a fixed element of the $d-$dimesional Euclidean space $\mathbb{R}^{d}$, none of the traditional risk measures $R(\hat{\mu}_{\HB},\mu)$ and $R(\hat{\mu}_{\PHB},\mu)$ are known to uniformly dominate the other. Hence, instead of assuming $\mu$ to be fixed, we consider $\mu$ to be generated according to a $N_{d}(0,\Sigma)$ distribution having a $d\times d$ positive definite covariance matrix $\Sigma$. Let $G_{0}$ be the probability measure associated with $N_{d}(0,\Sigma)$. Since $\mu$ is assumed to be random, for any estimator $\hat{\mu}$ of $\mu$, we consider the following integrated squared $\ell_2$ risk function of $\hat{\mu}$ given by
	\begin{equation}\label{RISK_FUNCTION}
		R(\hat{\mu},G_{0})=\int_{\mathbb{R}^d} E_{Y\mid \mu}\lVert \hat{\mu}-\mu\rVert^2_{2} dG_{0}(\mu).
	\end{equation}
	
	In the frequentist analysis of Bayesian methods, the parameter is regarded as a fixed unknown point, and the likelihood-prior combination yields a posterior distribution that serves as a measure-valued estimate. In contrast, the methodology employed in the present analysis assumes a generative distribution for the parameter and integrates it to define risk. This approach offers an alternative perspective on analyzing mixed-effects models, enabling a more nuanced evaluation of the efficacy of various procedures. Such an assumption of a `true prior' being part of the generative mechanism is common in the information theory and statistical physics literature; see, for example, Section 1.2.2 of the review article \cite{Zdeborova_2016}.
	
	Our study focuses on the comparison of the risk functions $R(\hat{\mu}_{\HB}, G_{0})$ and $R(\hat{\mu}_{\PHB}, G_{0})$ in a non--asymptotic framework, and identifying conditions on $\Sigma$ under which $R(\hat{\mu}_{\HB}, G_{0})<R(\hat{\mu}_{\PHB}, G_{0})$. Intuitively, due to the presence of the shared hyperparameter $\eta$ in model (\ref{HB_MODEL}), the hierarchical Bayes model should be able to leverage more information about $\mu$. Therefore, on an average, model (\ref{HB_MODEL}) should yield more accurate inference about $\mu$ compared to model (\ref{PHB_MODEL}). However, when the random effects $\mu_j$'s are independent or weakly correlated, model (\ref{PHB_MODEL}) is a more accurate approximation of the true $\mu-$distribution $G_{0}$ than model (\ref{HB_MODEL}). Thus, in such cases, $\hat{\mu}_{\PHB}$ should be a better choice in terms of having a smaller integrated risk compared to $\hat{\mu}_{\HB}$.
    
	\section{Integrated Risk Comparisons}
	In this section, we present the main theoretical results of this paper under the setup of Section 2. Our aim here is to study the integrated risk profiles $R(\hat{\mu}_{\HB}, G_{0})$ and $R(\hat{\mu}_{\PHB}, G_{0})$, and find conditions on the true $\mu-$generating mechanism $G_{0}$ under which $\hat{\mu}_{\HB}$ offsets $\hat{\mu}_{\PHB}$ by borrowing information through the shared hyperparameter $\eta$. Proposition \ref{PROP_INTRISK_COMPND_SYMM} presents explicit analytic forms of the integrated risks $R(\hat{\mu}_{\HB}, G_{0})$ and $R(\hat{\mu}_{\PHB}, G_{0})$ when $\Sigma=\Sigma_{d}(\rho)$ and $\Sigma_{d}(\rho)$ denotes the compound symmetric correlation matrix $(1-\rho) I_{d}+\rho 1_{d}1_{d}^{\T}$ with $-1/(d-1)<\rho<1$. Throughout this paper, $1_{d}$ denotes a $d\times1$ vector whose components are all equal to $1$. Analysis of the difference $R(\hat{\mu}_{\PHB}, G_{0})-R(\hat{\mu}_{\PHB}, G_{0})$ when $\Sigma=\Sigma_{d}(\rho)$ is given in Theorem \ref{THM_RISK_COMP_COMPND_SYMM}. Proposition \ref{PROP_INTRISK_GEN} generalizes the result of Proposition \ref{PROP_INTRISK_COMPND_SYMM} and gives explicit analytic expressions of $R(\hat{\mu}_{\HB},G_{0})$ and $R(\hat{\mu}_{\PHB},G_{0})$ when $\Sigma$ is arbitrary. In Theorem \ref{THM_RISK_COMPARISON_GENERAL}, we consider a broad class of $\mu-$generating mechanism $G_{0}$ obtained by perturbing the largest eigenvalue of $\Sigma_{d}(\rho)$ and analyze conditions under which $\hat{\mu}_{\HB}$ outperforms $\hat{\mu}_{\PHB}$. Proofs of these results along with those of propositions \ref{PROP_INTRISK_COMPND_SYMM} and \ref{PROP_INTRISK_GEN} can be found in the Appendix. Proofs of both theorems \ref{THM_RISK_COMP_COMPND_SYMM} and \ref{THM_RISK_COMPARISON_GENERAL} are quite long and tedious. Hence, we also include brief sketches of the proofs of both theorems in the Appendix before presenting their detailed analytical proofs.
	\begin{proposition}\label{PROP_INTRISK_COMPND_SYMM}
		Consider the random effects model (\ref{TRUE_MODEL}), where $\mu\sim N_{d}(0,\Sigma_{d}(\rho))$ with $-1/(d-1)<\rho<1$, $d\geq 5$ and $n\ge 1$. Then the integrated risk functions of $\hat{\mu}_{\HB}$ and $\hat{\mu}_{\PHB}$ are respectively given by\small
		\begin{eqnarray}
			R(\hat{\mu}_{\HB},\rho)&=&\frac{d}{n}-\frac{d-3}{n}\frac{1}{1+n(1-\rho)},\quad -\frac{1}{d-1}<\rho<1, \label{RISK_HB}\\
			R(\hat{\mu}_{\PHB},\rho)&=&\frac{d}{n}-\frac{(d-2)^2}{2n}\int_{0}^{1} \frac{u^{\frac{d-2}{2}-1}}{\surd{\xi(u;\rho)}}du,\quad -\frac{1}{d-1}<\rho<1, \label{RISK_PHB}
		\end{eqnarray}\normalsize
		where $\xi(u;\rho)=\big[1+n\{1+(d-1)\rho\}(1-u)\big]\big[1+n(1-\rho)(1-u)\big]^{d-1}$, $u\in[0,1]$. 
	\end{proposition}
	Clearly, $\max\{R(\hat{\mu}_{\HB},\rho), R(\hat{\mu}_{\PHB},\rho)\}< d/n=R((\bar{Y}_{1},\dots,\bar{Y}_{d}),\rho)$ for all $-1/(d-1)<\rho<1$, since the second terms on the right-hand sides of each of (\ref{RISK_HB}) and (\ref{RISK_PHB}) are strictly positive. Here, $(\bar{Y}_{1},\dots,\bar{Y}_{d})$ is the frequentist maximum likelihood estimator of $\mu$ that can be obtained as a Bayes estimator by assigning a uniform non-informative prior to $\mu$. This is in concordance with the celebrated James-Stein shrinkage effect for the joint estimation of $\mu$. When $\rho\approx 1$, that is, $\mu_{j}$s are strongly correlated among themselves, $R(\hat{\mu}_{\HB},\rho)\approx 3/n$ which is independent of $d$. In such a case, learning about an individual $\mu_{j}$ is enough to learn about all of them jointly. Also, for a fixed $d$, if a large number of replications corresponding to each $\mu_{j}$ is available, both models (\ref{PHB_MODEL}) and (\ref{HB_MODEL}) should borrow an equal amount of information from all the $Y$'s. Hence, $R(\hat{\mu}_{\PHB},\rho)\approx R(\hat{\mu}_{\HB},\rho)$ as $n\rightarrow\infty$, provided $d$ is fixed. Similar conclusions hold for $\hat{\mu}_{\PHB}$.\par
	
	Various studies were reported in the literature where researchers sought to discover the exact analytic expressions of classical risk measures for the James-Stein type estimators of the form (\ref{PHB_ESTIMATES}) and (\ref{HB_ESTIMATES}) \citep{Bock_1975, Egerton_Laycock_1982}. Unfortunately, these expressions often involve infinite series or complicated functions, such as Dawson's integrals, which can be challenging to manage. Thus, classical risk measures are not very useful in our particular context.\par
	Before proceeding to Theorem \ref{THM_RISK_COMP_COMPND_SYMM}, let us first define certain important functions and real quantities that would be essential to describe Theorem \ref{THM_RISK_COMP_COMPND_SYMM} and are presented in the display (\ref{Common_Display}) below. Of them, the quantities $\alpha_{d}$ and $\delta_{\ast}$ would be useful for describing Theorem \ref{THM_RISK_COMPARISON_GENERAL} which can be seen later as we proceed further.\par
	
	For $\rho\in\big(-1/(d-1),1\big)$, define the functions $ h_{\rho}(u)$ and $f_{\rho}(u)$ on the unit interval $[0,1]$ as
	\small\begin{align}\label{Common_Display}
		\begin{aligned}
			h_{\rho}(u)&=\bigg[\frac{1+n(1-\rho)(1-u)}{1+n(1+(d-1)\rho)(1-u)}\bigg]^{1/2},\quad f_{\rho}(u)=\frac{(d-2)[1+n(1-\rho)]u^{(d-2)/2-1}}{2\big[1+n(1-\rho)(1-u)\big]^{d/2}},\\
			I(\rho)&=\int_{0}^{1}h_{\rho}(u)f_{\rho}(u)du,\quad 
			\delta_{\ast}=\frac{1}{dn}\bigg[\frac{1}{\lim\limits_{\rho\uparrow 1} I^2(\rho)}-1 \bigg],\\
			\alpha_{d}&=\frac{\big(d-3\big)^2}{\big(d-2\big)^2}, \quad
			\rho_{U}=\frac{(1-\alpha_{d})(1+\frac{1}{n\delta_{\ast}})}{1+(d-1)\alpha_{d}},\quad \rho_{L}=\frac{(1-\alpha_{d})(1+\frac{1}{n})}{1+(d-1)\alpha_{d}}.
		\end{aligned}
	\end{align}\normalsize
	Some important remarks regarding the functions and quantities defined in (\ref{Common_Display}) are as follows. First, for each fixed $\rho$, $h_{\rho}(u)\in (0,1)$ and $f_{\rho}(u)\geq 0$ for all $u\in[0,1]$, and $\int_{0}^{1}f_{\rho}(u)du=1$. Moreover, for each fixed $d\ge 5$ and every $n\ge1$, $I(\rho)$ strictly lies in $(0,1)$ for all $\rho$ and $\lim_{\rho\uparrow 1} I(\rho)$ is bounded away from $1$ which ensures $\delta_{\ast}\in(0,1)$. Also, $0<\rho_{L}<\rho_{U}<1$ for each fixed $d\ge 5$ and every fixed $n\ge1$. We refer the interested readers to the proof of Theorem \ref{THM_RISK_COMP_COMPND_SYMM} to verify these assertions.
	\begin{theorem}\label{THM_RISK_COMP_COMPND_SYMM}
		Consider the setup of Proposition \ref{PROP_INTRISK_COMPND_SYMM}. Let $H(\rho)$ denote the difference between the integrated risks (\ref{RISK_PHB}) and (\ref{RISK_HB}). Then there exists a unique $\rho^{\ast}\in(\rho_{L},\rho_{U})$ such that
		\small  \begin{equation*}\label{RISK_DIFF_CONDITION}
			\mathop{H(\rho)} \left\{
			\begin{array}{ll}
				< 0 & \mbox{ if}\; -\frac{1}{d-1}< \rho < \rho^{\ast},\\ \\
				= 0 & \mbox{ if}\; \rho = \rho^{\ast},\\ \\
				> 0 & \mbox{ if}\; \rho^{\ast} <\rho <1.
			\end{array}\right.
		\end{equation*}\normalsize
		Moreover, the function $H(\rho)$ is strictly increasing in $\rho$ over $(0,1)$.
	\end{theorem}
	
	Theorem \ref{THM_RISK_COMP_COMPND_SYMM} therefore says that $\hat{\mu}_{\HB}$ yeilds an improved inference on average compared to $\hat{\mu}_{\PHB}$ whenever $\rho>\rho^{\ast}$. Using the arguments used to prove Theorem \ref{THM_RISK_COMP_COMPND_SYMM}, we obtain $\rho^{\ast}\sim(1-\alpha_{d})/\{1+(d-1)\alpha_{d}\}\sim 1/(d-1)$, provided $d$ is large and $n\ge1$ is held fixed. Our simulation study in Section 4 also suggests that $\rho^{\ast}\sim  1/d\approx 1/(d-1)$ is close to zero even if $d$ is moderately large. In most of the practical applications of the model (\ref{TRUE_MODEL}), we typically encounter moderate to large values of $d$. Consequently, $H(\rho)>0$ for almost all positive values of $\rho$. On the other hand, when $\rho$ lies in a small neighborhood around zero whose size is of the order of $1/(d-1)\approx 0$, model (\ref{PHB_MODEL}) provides a better approximation to $G_{0}$. In such situations, the proximity of the prior distribution to the true state of nature overrides the effect of information borrowing. Under such circumstances, model (\ref{PHB_MODEL}) yields better performance than model (\ref{HB_MODEL}). However, such situations are relatively rare in practice. Thus, barring some rare occasions, $\hat{\mu}_{\HB}$ always outperforms $\hat{\mu}_{\PHB}$ in the sense of having a smaller risk profile. This should be attributed to the presence of the shared component $\eta$ in the model (\ref{HB_MODEL}), which helps leverage more information from all the sample points.
	\begin{proposition}\label{PROP_INTRISK_GEN}
		Consider the random effects model (\ref{TRUE_MODEL}), where $\mu\sim N_{d}(0,\Sigma)$, $\Sigma$ being any arbitrary $d\times d$ positive definite covariance matrix with $d\geq 4$. Then the integrated risk function $R(\hat{\mu}_{HB},G_{0})$ of $\hat{\mu}_{\HB}$ is given by\small
		\begin{equation}\label{RISK_HB_GENERAL_CASE}
			R(\hat{\mu}_{\HB},G_{0})=\frac{d}{n}-\frac{(d-3)^2}{2n}\int_{0}^{1}\frac{u^{(d-3)/2-1}}{\surd{\xi_{1}(u)}}du,
		\end{equation}\normalsize
		while the integrated risk function $R(\hat{\mu}_{\PHB},G_{0})$ of $\hat{\mu}_{\PHB}$ is\small
		\begin{equation}\label{RISK_PHB_GENERAL_CASE}
			R(\hat{\mu}_{\PHB},G_{0})=\frac{d}{n}-\frac{(d-2)^2}{2n}\int_{0}^{1}\frac{u^{(d-2)/2-1}}{\surd{\xi_{2}(u)}}du,
		\end{equation} \normalsize
		where for $u\in[0,1]$ \small
		\begin{eqnarray}
			\xi_{1}(u)&=&\bigg[1-\frac{n(1-u)}{d}\sum_{j=1}^{d}\frac{\lambda_{j}Z_{j}^2}{1+n(1-u)\lambda_{j}} \bigg]\prod_{j=1}^{d}\big\{1+n(1-u)\lambda_{j}\big\},\label{RISK_HB_GENERAL_CASE_1}\\
			\xi_{2}(u)&=&\prod_{j=1}^{d}\big\{1+n(1-u)\lambda_{j}\big\},\quad u\in[0,1],\label{RISK_PHB_GENERAL_CASE_1}
		\end{eqnarray}\normalsize
		and $\lambda_{1},\dots,\lambda_{d}>0$ denote the eigenvalues of $\Sigma$. Here, $Z=P^{\T}1_{d}$, where $P$ is a $d\times d$ orthogonal matrix such that $P\Lambda P^{\T}=\Sigma$ and $\Lambda$ is a $d\times d$ diagonal matrix having diagonal elements $\lambda_{1},\dots,\lambda_{d}$.
	\end{proposition}
	
	Using Proposition \ref{PROP_INTRISK_GEN}, $R(\hat{\mu}_{\PHB},\rho)$ in (\ref{RISK_PHB}) can be immediately deduced as a special case of (\ref{RISK_PHB_GENERAL_CASE}) when $\Sigma=\Sigma_{d}(\rho)$. However, the same is not true for $R(\hat{\mu}_{\HB},\rho)$ in (\ref{RISK_HB}) as it requires some additional work based on the proof of Proposition \ref{PROP_INTRISK_COMPND_SYMM}. Here also we observe the Stein--shrinkage effect since $\max\{R(\hat{\mu}_{\HB},G_{0}), R(\hat{\mu}_{\PHB},G_{0})\}< d/n=R((\bar{Y}_{1},\dots,\bar{Y}_{d}),G_{0})$ for any arbitrary $\Sigma$.\par
	
	Theorem \ref{THM_RISK_COMPARISON_GENERAL} presents conditions under which $\hat{\mu}_{\HB}$ has a smaller integrated risk profile compared to that of $\hat{\mu}_{\PHB}$ within a general framework. We consider here a broad family of true $\mu-$generating distributions $G_{0}$ obtained by perturbing the largest eigenvalue of $\Sigma_{d}(\rho)$ while keeping the rest unaltered. Before stating Theorem \ref{THM_RISK_COMPARISON_GENERAL}, let us now describe certain important quantities required to describe the result.\par
	Define \small
	\begin{align}\label{Common_Display_2}
		\begin{aligned}
			B&=\frac{1}{n}[\{\alpha_{d}(1+nd\delta_{\ast})\}-1]>0, \quad
			\beta_{\ast}=\frac{1}{\surd{(1+n\nu)}},\ 0< \nu <B,\\
			\alpha^{\ast}_{d}&=\beta_{\ast}^2\alpha_{d},\quad
			\widetilde{\rho}_{U}=\frac{(1-\alpha^{\ast}_{d})(1+\frac{1}{n\delta_{\ast}})}{1+(d-1)\alpha^{\ast}_{d}},\quad
			\widetilde{\rho}_{L}=\frac{(1-\alpha^{\ast}_{d})(1+\frac{1}{n})}{1+(d-1)\alpha^{\ast}_{d}},
		\end{aligned}
	\end{align}\normalsize
	where the quantities $\alpha_{d}$ and $\delta_{\ast}$ have been defined already in (\ref{Common_Display}). The arguments used in the proof of Theorem \ref{THM_RISK_COMP_COMPND_SYMM} ensure that $B>0$ for each fixed $d\ge5$ and every $n\ge 1$. Again, it can be shown that both the constants $\widetilde{\rho}_{U}$ and $\widetilde{\rho}_{L}$ strictly lie inside the unit interval $(0,1)$ with $\widetilde{\rho}_{L}<\widetilde{\rho}_{U}$. Interested readers are referred to the proofs of Theorems 1 and 2 to validate these assertions.
	\begin{theorem}\label{THM_RISK_COMPARISON_GENERAL}
		Consider the setup of Proposition \ref{PROP_INTRISK_GEN}. Assume $\lambda_{j}=\lambda_{0,j}$, for $j=1,\dots,d-1$, and $\lambda_{d}=\lambda_{0,d}+\nu$, where $0<\lambda_{0,1}=\dots=\lambda_{0,d-1}<\lambda_{0,d}$ denote the eigenvalues of $\Sigma_{d}(\rho)$ as in Theorem \ref{THM_RISK_COMP_COMPND_SYMM}, and $0<\nu<B$, where $B$ is defined in (\ref{Common_Display_2}). Further assume $P=P_{0}$, where $P_{0}$ is a $d\times d$ orthogonal matrix such that $P_{0}\Lambda_{0} P_{0}^{\T}=\Sigma_{0}$ and $\Lambda_{0}$ is a $d\times d$ diagonal matrix having diagonal entries $\lambda_{0,1},\dots,\lambda_{0,d}$. Then there exists a unique $\rho^{\ast}\in(\widetilde{\rho}_{L},\widetilde{\rho}_{U})$ such that for all $ \rho^{\ast} <\rho <1$\small
		\begin{equation}\label{PERTURB_RISK_DIFF_CONDITION}
			R(\hat{\mu}_{\HB},G_{0}) < R(\hat{\mu}_{\PHB},G_{0}).\nonumber
		\end{equation}\normalsize
	\end{theorem}
	The main technical hurdle in analyzing the difference $R(\hat{\mu}_{\HB}, G_{0})-R(\hat{\mu}_{\PHB}, G_{0})$ for a general $\Sigma$ arises from the fact that $\xi_{1}(u)$ cannot be expressed in a simple analytic form due to the presence of $Z_{1},\dots, Z_{d}$ and $\lambda_{1},\dots,\lambda_{d}$, which can be quite arbitrary. Hence, in Theorem \ref{THM_RISK_COMPARISON_GENERAL}, we confine our attention to the case $\lambda_{j}=\lambda_{0,j}=1-\rho$ for $j=1,\dots,d-1$, and $\lambda_{d}=\lambda_{0,d}+\nu=1+(d-1)\rho+\nu$ with $0<\nu<B$. Since $P=P_{0}$, we have $Z=P^{\T}1_{d}=P_{0}^{\T}1_{d}$. Hence, the $Z_{j}$'s here would be the same as those in Theorem \ref{THM_RISK_COMP_COMPND_SYMM}. This helps to exploit the architecture of the proof of Theorem \ref{THM_RISK_COMP_COMPND_SYMM} and utilize certain facts already established there.\par
	
	As before, one can conclude that $\rho^{\ast}$ in Theorem \ref{THM_RISK_COMPARISON_GENERAL} is of the order of $(1-\alpha^{\ast}_{d})/\{1+(d-1)\alpha^{\ast}_{d}\}\sim 1/(d-1)\approx 0$, provided $d$ is large and $n\ge1$ is held fixed. In many practical applications of the model (\ref{TRUE_MODEL}), $d$ is usually moderately large or large. Hence, barring some rare cases, $\hat{\mu}_{\HB}$ should demonstrate a superior performance compared to $\hat{\mu}_{\PHB}$ on average, provided $\rho>0$. This should be attributed to the leverage of more information through the additional level of hierarchy for $\eta$ in the model (\ref{HB_MODEL}). Moreover, when $d$ is fixed, and $n$ is large, $B\approx0$, whence the amount of perturbation $\nu$ becomes infinitesimally small. In that case, $R(\hat{\mu}_{\HB},G_{0}) \approx R(\hat{\mu}_{\PHB}, G_{0})$ as $n\rightarrow\infty$ and the effect of borrowing of strength can be offset by drawing information from a sufficiently large sample.
	\section{Numerical Study}
\begin{figure}[H]
		\centering
		\begin{subfigure}{.42\textwidth}
			\centering
			\includegraphics[scale=0.4]{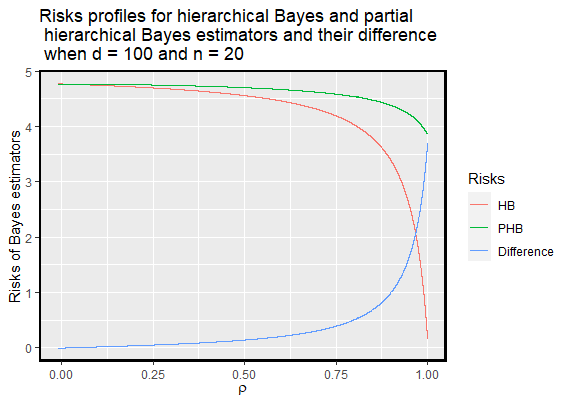}
			\caption{Plot showing risk functions and their difference}
			\label{Risk_Plot}
		\end{subfigure}
		\begin{subfigure}{.4\textwidth}
			\centering
			\includegraphics[scale=0.42]{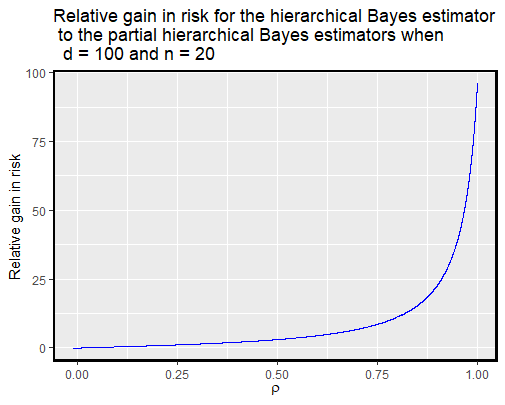}
			\caption{Plot showing relative gain in risk}
			\label{Gain_Risk_Plot}
		\end{subfigure}
		\caption{A figure with two subfigures showing the integrated risk profiles of the hierarchical and the partial-hierarchical Bayes estimators, the difference between these two risk functions, and the relative gain in risk for the hierarchical Bayes estimator when $d=100$, $n=20$ and $\Sigma=\Sigma_{d}(\rho)$.}
	\end{figure}

	In this section, we present a simulation study to numerically investigate the properties of $R(\hat{\mu}_{\HB}, \rho)$, $R(\hat{\mu}_{\PHB}, \rho)$ and their difference $H(\rho)$ when $\mu\sim N_{d}(0,\Sigma_{d}(\rho))$ with $-1/(d-1)<\rho<1$. Figure \ref{Risk_Plot} presents a plot of these three functions when $d=100$ and $n=20$. The explicit analytic forms of $R(\hat{\mu}_{\HB}, \rho)$ and $R(\hat{\mu}_{\PHB}, \rho)$ are given by (\ref{RISK_HB}) and (\ref{RISK_PHB}), respectively. Figure \ref{Gain_Risk_Plot} presents the plot of the integrated risk function $(R(\hat{\mu}_{\HB}, \rho) - R(\hat{\mu}_{\PHB}, \rho))/R(\hat{\mu}_{\PHB}, \rho)\times 100$ across the $\rho$ values. The quantity aforementioned is a measure of the relative gain in risk from $\hat{\mu}_{\HB}$ to $\hat{\mu}_{\PHB}$. It is expressed in percentages and hence is unit-free.
	
Both Figure \ref{Risk_Plot} and Figure \ref{Gain_Risk_Plot} show that for values of $\rho$ close to zero, that is, when $\mu_{j}$s are weakly correlated, $\hat{\mu}_{\PHB}$ performs slightly better than $\hat{\mu}_{\HB}$ in terms of having a smaller risk. This is not surprising since under such a weak correlation structure, model (\ref{PHB_MODEL}) provides a closer approximation to the true joint distribution of the $\mu_{j}$s than model (\ref{HB_MODEL}). However, as $\rho$ increases, both the difference between risks and the relative gain in risk for $\hat{\mu}_{\HB}$ increase, eventually become positive, and increase steadily thereafter along the increasing $\rho$ values. Let $\rho^{\ast}$ be the unique root of the difference $H(\rho)$. Clearly, when $\rho>\rho^{\ast}$, $R(\hat{\mu}_{\HB}, \rho)<R(\hat{\mu}_{\PHB}, \rho)$ and, therefore, $\hat{\mu}_{\HB}$ outperforms $\hat{\mu}_{\PHB}$ for all such values of $\rho$. In this setup, we numerically estimate $\rho^{\ast}$ to be approximately $0.0631$. 


We also estimate the regression equation of the logarithm of $\rho^{\ast}$ on the logarithm $\log(d)$ of the dimension $d$ and the logarithm $\log(n)$ of the common sample size $n$ based on $500$ different combinations of $(d,n)$. The aforesaid estimated regression function is found to be $\log(\rho^{\ast})= 1.27353-1.18846*\log(d)$ with an adjusted $R^2$ of $0.9941$. Here, the estimated regression coefficient of $\log(d)$ is close to $1$, and $\log(n)$ is found to be statistically insignificant for the aforementioned regression model. Thus, our simulation study indicates that $\rho^{\ast}$ should be approximately of the order of $1/d$. This is theoretically confirmed in Section 3 of this paper. Thus, except for the small region $(-1/(d-1),\rho^{\ast})=(-0.0526,0.0631)$ whose size is of the order of $1/(d-1)$, $\hat{\mu}_{\HB}$ is always preferable to $\hat{\mu}_{\PHB}$ in terms of having a smaller risk. The gain in risk for $\hat{\mu}_{\HB}$ is attributed to the borrowing of strength by $\eta$ in model (\ref{HB_MODEL}), which is absent in model (\ref{PHB_MODEL}).
\section{Discussion}
In this article, we conducted a rigorous analytical exploration of the role that shared hyperparameters play in hierarchical Bayesian models, highlighting their crucial impact on information borrowing and posterior inference.  By introducing an integrated risk measure relative to the true data-generating distribution, we were able to compare
estimators arising from models with differing levels of hierarchy. Our analysis, while specific to a one-group hierarchical Gaussian model, suggests that the benefits of hierarchical Bayes do not stem merely from intuition or empirical success, but can be rigorously formalized through an integrated risk framework. Our analysis shows that additional layers of hierarchy translate into measurable efficiency gains whenever the underlying random effects exhibit even modest correlation. This reframes the notion of “borrowing strength” from a heuristic into a quantifiable principle with precise analytical conditions.

These findings have repercussions well beyond the normal-means model. In practice, settings such as multi-center clinical trials, meta-analyses, and high-dimensional genomic studies routinely involve correlated groups where borrowing is essential. In such cases,
the integrated risk may function as a practical ``borrowing index,'' offering practitioners a direct and interpretable way to assess how much information is shared across groups. Similarly, for more complex hierarchical structures—such as generalized linear mixed
models or Bayesian nonparametrics—our results suggest that deeper hierarchies should generally outperform shallower ones whenever shared hyperparameters capture genuine dependence across units.

While a deeper hierarchical structure can offer certain advantages, it comes at an expense of elevated computing costs. Each additional layer introduces hyperparameters that may be weakly identified, raising challenges of prior sensitivity, slower mixing of MCMC algorithms, and heavier computational burden. This trade-off
suggests the need for adaptive strategies that balance the statistical benefits of borrowing with the risks of inefficiency. One promising avenue is to develop data-driven estimators of the integrated risk, which could support empirical selection of hierarchical depth and hyperparameters. Such an approach would align naturally with modern predictive evaluation tools such as W-AIC and cross-validation \citep{Vehtari_etal_2017, Watanabe_2010}, and it could help connect our framework to optimal shrinkage theory, where borrowing is tuned to achieve minimax efficiency under structural constraints.

Taken together, these insights suggest that hierarchical Bayes is not a one-size-fits-all solution, but a flexible modeling strategy whose depth should be calibrated to the specific scientific context and computational resources available. By making explicit the conditions under which deeper hierarchies pay off, our integrated risk perspective provides a principled basis for that calibration and lays the groundwork for future work on adaptive and computationally efficient hierarchical modeling.

\section*{Acknowledgement}
Dr. Bhattacharya and Dr. Pati acknowledge NSF DMS-2413715 for partially supporting this research.



\appendices
\section*{Appendix}

In this section, we present the proofs of the main theoretical results of the paper, namely, Theorem \ref{THM_RISK_COMP_COMPND_SYMM} and Theorem \ref{THM_RISK_COMPARISON_GENERAL}. However, proofs of these results are quite long and tedious. Hence, before presenting the detailed analytical proofs, let us first outline the proofs of both theorems briefly.
\subsection*{\textbf{Sketch of the proof of Theorem \ref{THM_RISK_COMP_COMPND_SYMM}:}}
\noindent In the following, we present a brief outline of the proof of Theorem \ref{THM_RISK_COMP_COMPND_SYMM} in a step-by-step manner. However, we only describe the main technical steps as follows.
\begin{enumerate}
\item By definition, the difference $H(\rho)$ of the integrated risks $R(\hat{\mu}_{\PHB},\rho)$ and $R(\hat{\mu}_{\HB},\rho)$ is a non--linear function in $\rho$ and it involves the integral $$\int_{0}^{1}u^{(d-2)/2-1}/\surd{\xi(u;\rho)}du$$ that cannot be evaluated explicitly in a simple analytic form. Hence, using the functions $h_{\rho}(u)$ and $f_{\rho}(u)$ defined in (\ref{Common_Display}), let us rewrite the function $H(\rho)$ as
\begin{equation}\label{H_rho_eqiuv_1_sktch}
	H(\rho) = \frac{\big[(d-3)-(d-2)\int_{0}^{1}h_{\rho}(u)f_{\rho}(u)du\big]}{n[1+n(1-\rho)]}.
\end{equation}

\item Using the representation (\ref{H_rho_eqiuv_1_sktch}) above and noting the fact that $\int_{0}^{1}f_{\rho}(u)du=1$, we have 
\begin{equation}\label{H_rho_eqiuv_2_sktch}
	H(\rho)=\frac{d-2}{n[1+n(1-\rho)]}\int_{0}^{1}(\surd{\alpha_{d}}-h_{\rho}(u))f_{\rho}(u)du,
\end{equation}
where $\alpha_{d}$ is defined in (\ref{Common_Display}). Since $h_{\rho}(u)>\surd{\alpha_{d}}$ for all $u\in(0,1)$ whenever $\rho\le\rho_{L}$, it follows from (\ref{H_rho_eqiuv_2_sktch}) that $H(\rho)<0$ for all $-1/(d-1)<\rho\leq\rho_{L}$. Here $\rho_{L}$ is already defined in (\ref{Common_Display}).
\item Next consider the case $\rho>0$. Using the function $I(\rho)$ defined in (\ref{Common_Display}), we note that
\begin{equation}
	H(\rho)=\frac{1}{n[1+n(1-\rho)]}\big[(d-3)-(d-2)I(\rho)].\label{H_rho_eqiuv_3_sktch}
\end{equation}
We would aim to first find some $\rho>0$ such that $H(\rho)>0$. Towards that, we show that $\lim_{\rho\uparrow 1}H(\rho)>0$ which is equivalent to showing $\lim_{\rho\uparrow 1}I(\rho)< \surd{\alpha_{d}}$, where
\begin{equation}\label{limit_I_rho_sktch}
	\lim\limits_{\rho\uparrow 1} I(\rho)=\int_{0}^{1} \frac{(d-2)u^{\frac{d-2}{2}-1}}{2\sqrt{1+nd(1-u)}}du.
\end{equation}
Since the integral on the right hand side of (\ref{limit_I_rho_sktch}) is strictly decreasing in $n$, we show $\lim_{\rho\uparrow 1}I(\rho)< \surd{\alpha_{d}}$ when $n=1$. This establishes that $\lim_{\rho\uparrow 1}H(\rho)>0$ for all $d\ge 5$ and for each $ n\ge 1$.

\item Using Step (3) we have $H(\rho)>0$ for each $\rho$ in a left neighborhood of $1$. So, by the mean value theorem for continuous functions, there must exist some $\rho^{\ast}\in (\rho_{L},1)$ such that $H(\rho^{\ast})=0$. However, this alone does not ensure the uniqueness of the root $\rho^{\ast}$. For that, we need to show $H(\rho)$ is strictly increasing in $\rho$ over $(0,1)$. Towards that, we first observe that the function $\rho\mapsto [1+n(1-\rho)]^{-1}$ is strictly increasing in $\rho$ over $(0,1)$. Therefore, in view of (\ref{H_rho_eqiuv_3_sktch}), it would be enough to show that $I(\rho)$ strictly decreases in $\rho\in(0,1)$. For that, let us first rewrite the function $I(\rho)$ as
\begin{equation}\label{I_rho_alt_sktch}
	I(\rho)=1-\int_{0}^{1}\bigg[\frac{u}{1+n(1-\rho)(1-u)}\bigg]^{\frac{d-2}{2}}\frac{\partial}{\partial u} h_{\rho}(u)du.
\end{equation}
Applying Leibnitz's theorem and noting that $\partial^2 h_{\rho}(u)/\partial \rho \partial u  > 0$ for each fixed $u\in(0,1)$, we show that the derived function of $I(\rho)$ is always strictly negative for all $\rho>0$.     Hence, $H(\rho)$ is strictly increasing in $\rho$ and changes it sign exactly once at $\rho=\rho^{\ast}$. This shows that $\rho^{\ast}$ is unique.


%


\item We next obtain some non--trivial $\rho_{U}\in(\rho_{L},1)$ such that $H(\rho_{U})>0$ as follows.
\begin{enumerate}[]
	\item Using the generalized mean value theorem for Riemann integrals and the fact $\int_{0}^{1}f_{\rho}(u)du=1$, it follows $I(\rho)= h_{\rho}(u_{0}(\rho))$ for some $u_{0}(\rho)\in[0,1]$. Since, for each fixed $\rho\in(0,1)$, $h_{\rho}(u)$ is continuous and strictly increases as a function of $u$, one must have $u_{0}(\rho)\in(0,1)$.
	\item Next using a contra--positive argument we show that $u_{0}(\rho)$'s are uniformly bounded away from $1$, that is, there exists some $\delta_{0}\in(0,1)$ which is independent of any $\rho\in(0,1)$, and which possibly depends on the dimension $d$ and the sample size $n$, such that $u_{0}(\rho) <1-\delta_{0}$ for all $\rho\in(0,1)$.
	\item Sine the function $\rho\mapsto h_{\rho}(u)$, $\rho\in(0,1)$ is strictly decreasing in $\rho$, inverse of $h_{\rho}$ exists and decreases strictly in $\rho$. Since the composition of two decreasing real-valued functions in one real variable is increasing, $u_{0}(\rho)=h_{\rho}^{-1}(I(\rho))$ is strictly increasing in $\rho$ over $(0,1)$.
	\item Again, $H(\rho)>0 \mbox{ if and only if } I(\rho)< \surd{\alpha_{d}},$
	that is, if and only if
	\begin{equation}\label{Condition_Positivity_2_sktch}
		\rho > \frac{1-\alpha_{d}}{1+(d-1)\alpha_{d}}\big[1+\frac{1}{n(1-u_{0}(\rho))}\big].
	\end{equation}
	Define $\delta_{\ast}=\inf_{\rho\in(0,1)}[1-u_{0}(\rho)]$ and consider the quantity $\rho_{U}$ defined in (\ref{Common_Display}). Using properties of the functions $h_{\rho}$, $I(\rho)$ and $u_{0}(\rho)$, we show that $\delta_{\ast}=(nd)^{-1}[1/\lim_{\rho\uparrow 1} I^2(\rho)-1]$ whence $\rho_{U}<1$ is ensured. Note that, in (\ref{Common_Display}), we only provide the deterministic value of $\delta_{\ast}$.
	
	\item Finally, we observe that
	\begin{equation}
		\rho_{U} = \sup_{\rho\in(0,1)}\frac{1-\alpha_{d}}{1+(d-1)\alpha_{d}}\bigg[1+\frac{1}{n(1-u_{0}(\rho))}\bigg].\nonumber
	\end{equation}
	Hence $H(\rho)>0$ whenever $\rho\ge\rho_{U}$ since all such $\rho$ satisfies (\ref{Condition_Positivity_2_sktch}). In particular, $H(\rho_{U})>0$. So using the mean value theorem once more, it follows $\rho^{\ast}\in (\rho_{L},\rho_{U})$ when we conclude our arguments.
\end{enumerate}
\end{enumerate}

\subsection*{\textbf{Sketch of the proof of Theorem \ref{THM_RISK_COMPARISON_GENERAL}:}}
Let us now present a brief outline of the proof of Theorem \ref{THM_RISK_COMPARISON_GENERAL} as follows.
\begin{enumerate}
\item Using Proposition \ref{PROP_INTRISK_GEN}, the risk function of the hierarchical Bayes estimator $\hat{\mu}_{\HB}$ is given by
\begin{equation}
	R(\hat{\mu}_{\HB},G_{0})=\frac{d}{n}-\frac{(d-3)^2}{2n}\int_{0}^{1}\frac{u^{(d-3)/2-1}}{\surd{\xi_{1}(u)}}du,\label{GEN_RISK_HB_sktch}
\end{equation}
where, for $u\in[0,1]$, $\xi_{1}(u)$ is defined as
\small
\begin{eqnarray}\label{GEN_XI_1_sktch}
	\xi_{1}(u)
	&=&\bigg[1-\frac{n(1-u)}{d}\sum_{j=1}^{d}\frac{\lambda_{j}Z_{j}^2}{1+n(1-u)\lambda_{j}} \bigg]\prod_{j=1}^{d}\big\{1+n(1-u)\lambda_{j}\big\}.
\end{eqnarray}
\normalsize

\item Note that
\begin{equation}\label{GEN_XI_1_UB2_sktch}
	\xi_{1}(u) 
	< \xi_{0,1}(u)\times \frac{1+n(1-u)(\lambda_{0,d}+\nu)}{1+n(1-u)\lambda_{0,d}},
\end{equation}
where
\begin{eqnarray}
	\xi_{0,1}(u)
	&=& \bigg[1-\frac{n(1-u)}{d}\sum_{j=1}^{d}\frac{\lambda_{0,j}Z_{j}^2}{1+n(1-u)\lambda_{0,j}} \bigg]\prod_{j=1}^{d}\big\{1+n(1-u)\lambda_{j}\big\}, \ u \in (0,1).\nonumber
\end{eqnarray}
\item Using Step (2), we show that 
\begin{eqnarray}\label{GEN_RISK_HB_UB1_sktch}
	R(\hat{\mu}_{\HB},G_{0})
	&<& \frac{d}{n}-\frac{(d-3)^2}{2n}\int_{0}^{1}\frac{u^{(d-3)/2-1}}{\surd{\xi_{0,1}(u)}}\times \psi_{\rho}(u)du,
\end{eqnarray}
where 
\begin{eqnarray}
	\psi_{\rho}(u)
	&=& \bigg[\frac{1+n(1-u)\{1+(d-1)\rho\}}{1+n(1-u)\{1+(d-1)\rho+\nu\}}\bigg]^{\frac{1}{2}}, \ u \in (0,1).\nonumber
\end{eqnarray}
\item Next we observe that, for each fixed $\rho$ $\psi_{\rho}(u)$ is strictly increasing in $u$ over $(0,1)$ whence
\begin{eqnarray}
	R(\hat{\mu}_{\HB},G_{0}) < \frac{d}{n}-\frac{d-3}{n[1+n(1-\rho)]}\zeta_{1}(\rho),\label{GEN_RISK_HB_UB3_sktch}
\end{eqnarray}
where $\zeta_{1}(\rho)=\inf_{u\in(0,1)}\psi_{\rho}(u)=[1+n\{1+(d-1)\rho\}]/[1+n\{1+(d-1)\rho+\nu\}]^{1/2}$.
\item Observe that the function $\rho\mapsto \zeta_{1}(\rho)$ is strictly increasing in $\rho$. Hence,
\begin{eqnarray}
	R(\hat{\mu}_{\HB},G_{0})
	&<& \frac{d}{n}-\beta_{\ast}\frac{d-3}{n[1+n(1-\rho)]},\label{GEN_RISK_HB_UB4_sktch}
\end{eqnarray}
where $\beta_{\ast}=\inf_{\rho}\zeta_{1}(\rho)=(1+n\nu)^{-1/2}$. Thus, we obtain an upper bound to $R(\hat{\mu}_{\HB},G_{0})$, a function that is linear in $(d-3)/[n\{1+n(1-\rho)\}]$ and having coefficient $-\beta_{\ast}\in(-1,0)$ with an intercept $d/n$. It should be noted that the risk function $R(\hat{\mu}_{\HB},\rho)$ as given in (\ref{RISK_HB}) has exactly the same linear form the only difference being the coefficient which is $-1$.

\item Again, the risk function of the partial--hierarchical Bayes estimator
$\hat{\mu}_{\PHB}$ can be bounded below as
\begin{eqnarray}
	R(\hat{\mu}_{\PHB},G_{0})
	&>& \frac{d}{n}-\frac{(d-2)^2}{2n}\int_{0}^{1} \frac{u^{\frac{d-2}{2}-1}}{\surd{\xi(u;\rho)}}du,\label{GEN_RISK_PHB_LB1_sktch}
\end{eqnarray}
\normalsize
where $\xi(u;\rho)$ is already defined in Proposition \ref{PROP_INTRISK_COMPND_SYMM}. Therefore, we obtain
\begin{eqnarray}
	R(\hat{\mu}_{\HB},G_{0})- R(\hat{\mu}_{\PHB},G_{0})
	&<& K(\rho)
\end{eqnarray}
\normalsize
where
\begin{equation}
	K(\rho)=\frac{(d-2)^2}{2n}\int_{0}^{1} \frac{u^{\frac{d-2}{2}-1}}{\surd{\xi(u;\rho)}}du-\beta_{\ast}\frac{d-3}{n[1+n(1-\rho)]}\label{GEN_RISK_DIFF_UB_sktch}
\end{equation}
\normalsize 
for $-1/(d-1)<\rho<1$. Comparing (\ref{GEN_RISK_DIFF_UB_sktch}) with $H(\rho)$, it is evident that the function $K(\rho)$ has exactly the same form as that of the negative of $H(\rho)$ except the presence of $-\beta_{\ast}\in(-1,0)$, the coefficient of the term $(d-3)/[n\{1+n(1-\rho)\}]$, which is $-1$ in case of $-H(\rho)$. Hence, adopting exactly the same line of arguments used to analyze the properties of $H(\rho)$ in Theorem \ref{THM_RISK_COMP_COMPND_SYMM}, one can establish similar theoretical properties of $K(\rho)$. This concludes the arguments of the proof of Theorem \ref{THM_RISK_COMPARISON_GENERAL}.
\end{enumerate}
In Theorem \ref{THM_RISK_COMPARISON_GENERAL}, we work with only a non--trivial upper bound to $R(\hat{\mu}_{\HB},G_{0})- R(\hat{\mu}_{\PHB},G_{0})$ whose analytic behavior is quite similar to that of $-H(\rho)$ as in Theorem \ref{THM_RISK_COMP_COMPND_SYMM}. This is because finding a non--trivial lower bound to $R(\hat{\mu}_{\HB},G_{0})$ that behaves similarly as that of $R(\hat{\mu}_{\HB},\rho)$ when $\Sigma=\Sigma_{d}(\rho)$, does not seem feasible. Consequently, we can only provide a range of $\rho$ values where $\hat{\mu}_{\HB}$ ofsets $\hat{\mu}_{\PHB}$ on an average.

\subsection*{\textbf{Proofs of the main theoretical results:}}
Now we present the proofs of Theorem \ref{THM_RISK_COMP_COMPND_SYMM} and Theorem \ref{THM_RISK_COMPARISON_GENERAL} of the main article. However, before going to the proofs, let us state and prove the following important lemma that will be instrumental in proving what follows next.
\begin{lemma}
\label{lemma_NCCS_expect}
Let $S$ be a real-valued random variable following a non-central $\chi^2$ distribution with $d$ degrees of freedom and non-centrality parameter $\lambda>0$. Then 
\begin{equation}
	E(S^{-1})=\frac{1}{2}\int_{0}^{1}u^{(d-2)/2-1}e^{-\lambda(1-u)/2}du.\nonumber
\end{equation}
\end{lemma}

\noindent{\textit{Proof:}}
First we observe that every non-central $\chi^2$ random variable $S$ with $d$ degrees of freedom and non-centrality parameter $\lambda>0$ can be represented equivalently as
\begin{align}\label{POISON_HIERARCHY}
	\begin{aligned}
		S\mid N &\sim \chi^{2}_{d+2N}\\
		N &\sim Poi \big(\lambda/2\big).
	\end{aligned}
\end{align}
Here $\chi^{2}_{a}$ denotes a central $\chi^{2}$ distribution with $a$ degrees of freedom and $Poi(\lambda)$ denotes a Poisson distribution with mean parameter $\lambda$. 

Next, we use a result from \cite{Egerton_Laycock_1982} that says that if $\pi_{X}(\cdot)$ denotes the probability generating function of a non-negative integer-valued random variable $X$ then
\begin{equation}
	\int_{0}^{1} \pi_{X}(z)dz= E\bigg(\frac{1}{X+1}\bigg).\nonumber
\end{equation}
Using the above result and the Poisson hierarchical representation (\ref{POISON_HIERARCHY}), it follows
\small \begin{eqnarray}
	E(S^{-1})
	= \int_{0}^{1} \pi_{2N+d-3}(z)dz 
	&=& \int_{0}^{1} z^{d-3}\pi_{N}(z^2)dz\nonumber\\
	&=& \int_{0}^{1} z^{d-3}e^{\lambda(z^2-1)/2}dz.\label{lemma_NCCS_expect_1}
\end{eqnarray}
\normalsize 
The last step in (\ref{lemma_NCCS_expect_1}) follows from the fact that the probability generating function $\pi_{K}$ of any Poisson random variable $K$ with mean $\kappa$ is given by $\pi_{K}(z)=e^{\kappa(z-1)/2}$. Now using the transformation $u=z^2$ in (\ref{lemma_NCCS_expect_1}), Lemma \ref{lemma_NCCS_expect} follows immediately.

\section*{Proof of Proposition \ref{PROP_INTRISK_COMPND_SYMM}:}

\noindent{\textit{Proof:}}
For each $j=1,\dots,d$, let us first rewrite $\hat{\mu}_{j,\HB}$ as
\begin{equation}\label{HB_ESTIMATES_ALT}
	\hat{\mu}_{j,\HB}=\bar{Y}_{j}-\frac{d-3}{n\sum_{j=1}^{d}(\bar{Y}_{j}-\bar{Y})^2} (\bar{Y}_{j}-\bar{Y}).
\end{equation}
Write $\bar{\mu}=d^{-1}\sum_{j=1}^{d}\mu_{j}$. Now, given $\mu\in\mathbb{R}^{d}$, $\bar{Y}_{1},\dots,\bar{Y}_{n}$ are independently distributed with $\bar{Y}_{j} \sim N(\mu_{j},n^{-1})$ for $j=1,\dots,d$. Hence, given $\mu\in\mathbb{R}^{d}$, $n\sum_{j=1}^{d}(\bar{Y}_{j}-\bar{Y})^{2}$ follows a non-central $\chi^2$ distribution with $d-1$ degrees of freedom and non-centrality parameter $n\sum_{j=1}^{d}(\mu_{j}-\bar{\mu})^2=n\mu^{\T}(I_{d}-d^{-1}1_{d}1_{d}^{\T})\mu$. Therefore, applying Stein's lemma \citep{Efron_2012, Stein_1981}, together with Lemma \ref{lemma_NCCS_expect}, it follows, for each $\mu\in \mathbb{R}^{d}$, one has
\small  \begin{eqnarray}\label{RISK_HB_1}
	E_{Y\mid\mu}\lVert \hat{\mu}_{\HB}-\mu\rVert^2_{2}
	&=& \frac{d}{n}-\bigg(\frac{d-3}{n}\bigg)^2 E_{Y\mid\mu}\frac{1}{\sum_{j=1}^{d}(\bar{Y}_{j}-\bar{Y})^{2}}\nonumber\\
	&=& \frac{d}{n}-\frac{(d-3)^2}{n}E_{Y\mid\mu}\frac{1}{n\sum_{j=1}^{d}(\bar{Y}_{j}-\bar{Y})^2}\nonumber\\
	&=& \frac{d}{n}-\frac{(d-3)^2}{2n}\int_{0}^{1}u^{(d-3)/2-1}I(\mu;u)du.
\end{eqnarray}
\normalsize
where $I(\mu; u)= e^{-n\mu^{\T}(I_{d}-d^{-1}1_{d}1_{d}^{\T})\mu(1-u)/2}$, $u\in[0,1]$ and $\mu\in\mathbb{R}^{d}$. Using (\ref{RISK_HB_1}) coupled with Fubini's theorem, it follows
\small \begin{eqnarray}\label{RISK_HB_2}
	R(\hat{\mu}_{\HB},\rho)
	&=& \int_{\mathbb{R}^d} E_{Y\mid\mu}\lVert \hat{\mu}_{\HB}-\mu\rVert^2_{2}dG_{0}(\mu)\nonumber\\
	&=& \frac{d}{n}-\frac{(d-3)^2}{2n}\int_{0}^{1}u^{(d-3)/2-1} \int_{\mathbb{R}^d} I(\mu; u) \phi_{\mu, \Sigma_{d}(\rho)}(0) d\mu du \nonumber\\ 
	&=& \frac{d}{n}-\frac{(d-3)^2}{2n}\int_{0}^{1}u^{(d-3)/2-1} \bigg\{\frac{|\Sigma_{\H}(u,\rho)|}{|\Sigma_{d}(\rho)|}\bigg\}^{1/2}du,
\end{eqnarray}
\normalsize
where $\Sigma_{\H}(u,\rho)=\big[n(1-u)(I_{d}-d^{-1}1_{d}1_{d}^{\T})+\Sigma_{d}(\rho)^{-1}\big]^{-1}$ is a $d\times d$ positive definite matrix for each $u\in[0,1]$. Now
\begin{eqnarray}\label{RISK_HB_3}
	\Sigma_{d}(\rho)\Sigma_{\H}(u,\rho)^{-1}
	&=&n(1-u)\Sigma_{d}(\rho)(I_{d}-d^{-1}1_{d}1_{d}^{\T})+I_{d}\nonumber\\
	&=& \big[1+n(1-\rho)(1-u)\big]I_{d} - d^{-1}n(1-\rho)(1-u)1_{d}1_{d}^{\T}\nonumber\\
	&=& A_{\H}+v_{\H}w_{\H}^{\T},
\end{eqnarray}
where $A_{\H}=\big[1+n(1-\rho)(1-u)\big]I_{d}$, $v_{\H}=-d^{-1}n\rho(1-u)1_{d}$ and $w_{\H}=1_{d}$.

Therefore, using (\ref{RISK_HB_3}) and the matrix determinant lemma, we obtain
\begin{eqnarray}\label{RISK_HB_4}
	\frac{|\Sigma_{d}(\rho)|}{|\Sigma_{\H}(u,\rho)|}
	&=& |\Sigma_{d}(\rho)\Sigma_{\H}(u,\rho)^{-1}|\nonumber\\
	&=& |A_{\H}|\big(1+w_{\H}^{\T}A_{\H}^{-1}v_{\H}\big)\nonumber\\
	&=& \big[1+n(1-\rho)(1-u)\big]^{d} \times \big[1-\frac{d^{-1}n\rho(1-u)}{1+n(1-\rho)(1-u)}1_{d}^{\T}1_{d}\big]\nonumber\\
	&=& \big[1+n(1-\rho)(1-u)\big]^{d-1}.
\end{eqnarray}
Combining (\ref{RISK_HB_2}) and (\ref{RISK_HB_4}) we finally obtain
\small \begin{eqnarray}\label{RISK_HB_5}
	R(\hat{\mu}_{\HB},\rho)
	&=& \frac{d}{n}-\frac{(d-3)^2}{2n}\int_{0}^{1}\frac{u^{(d-3)/2-1}}{\big[1+n(1-\rho)(1-u)\big]^{(d-1)/2}}du\nonumber\\
	&=& \frac{d}{n}-\frac{d-3}{n}\frac{1}{1+n(1-\rho)}.
\end{eqnarray}
\normalsize
The last step in (\ref{RISK_HB_5}) follows from the fact that, for any $r>0$,
\begin{equation}
	\frac{d}{du}\bigg(\frac{u}{1+n(1-\rho)(1-u)}\bigg)^{r/2}=\frac{r[1+n(1-\rho)]u^{r/2-1}}{2\big[1+n(1-\rho)(1-u)\big]^{r/2+1}},\nonumber
\end{equation}
whence
\begin{equation}
	\int_{0}^{1}\frac{ru^{r/2-1}}{2\big[1+n(1-\rho)(1-u)\big]^{r/2+1}}du=\frac{1}{1+n(1-\rho)}.\label{RISK_HB_6}
\end{equation}
This completes the proof of (\ref{RISK_HB}). 

As before, for each $\mu\in \mathbb{R}^{d}$, we obtain
\small  \begin{eqnarray}
	E_{Y\mid\mu}\lVert \hat{\mu}_{\PHB}-\mu\rVert^2_{2}
	&=& \frac{d}{n}-\bigg(\frac{d-2}{n}\bigg)^2E_{Y\mid\mu}\frac{1}{\sum_{j=1}^{d}\bar{Y}_{j}^{2}}\nonumber\\
	&=& \frac{d}{n}-\frac{(d-2)^2}{n}E_{Y\mid\mu}\frac{1}{n\sum_{j=1}^{d}\bar{Y}_{j}^{2}}\nonumber\\
	&=& \frac{d}{n}-\frac{(d-2)^2}{2n}\int_{0}^{1}u^{(d-2)/2-1}J(\mu; u)du,\quad \label{PHB_RISK_1}
\end{eqnarray}
\normalsize
where $J(\mu; u) = e^{-n\mu^{\T}\mu(1-u)/2}$, $u\in[0,1]$ and $\mu\in\mathbb{R}^{d}$. The last step in (\ref{PHB_RISK_1}) follows from the fact that, given $\mu\in \mathbb{R}^{d}$, $n\sum_{j=1}^{d}\bar{Y}_{j}^{2}$ follows a non-central $\chi^2$ distribution with $d$ degrees of freedom and non-centrality parameter $n\mu^{\T}\mu$ together with Lemma \ref{lemma_NCCS_expect}.

Using (\ref{PHB_RISK_1}) and applying Fubini's theorem as before, it follows
\begin{eqnarray}
	R(\hat{\mu}_{\PHB},\rho)
	&=& \int_{\mathbb{R}^d} E_{Y\mid\mu}\lVert \hat{\mu}_{\PHB}-\mu\rVert^2_{2}dG_{0}(\mu)\nonumber\\
	&=& \frac{d}{n}-\frac{(d-2)^2}{2n}\int_{0}^{1}u^{(d-2)/2-1} \int_{\mathbb{R}^d} J(\mu; u) \phi_{\mu, \Sigma_{d}(\rho)}(0) d\mu du \nonumber\\
	%
	&=& \frac{d}{n}-\frac{(d-2)^2}{2n}\int_{0}^{1}u^{(d-2)/2-1} \bigg\{\frac{|\Sigma_{\PH}(u,\rho)|}{|\Sigma_{d}(\rho)|}\bigg\}^{1/2}du,
	\label{PHB_RISK_2}
\end{eqnarray}
where $\Sigma_{\PH}(u,\rho)=\big[n(1-u)I_{d}+\Sigma_{d}(\rho)^{-1}\big]^{-1}$ is a $d\times d$ positive definite matrix for each $u\in[0,1]$. Next
\begin{eqnarray}
	\Sigma_{d}(\rho)\Sigma_{\PH}(u,\rho)^{-1}
	&=&n(1-u)\Sigma_{d}(\rho)+I_{d}\nonumber\\
	&=& \big[1+n(1-\rho)(1-u)\big]I_{d}+n\rho(1-u)1_{d}1_{d}^{\T}\nonumber\\
	&=& A_{\PH}+v_{\PH}w_{\PH}^{\T}\label{interm} 
\end{eqnarray}
where $A_{\PH}=\big[1+n(1-\rho)(1-u)\big]I_{d}$, $v_{\PH}=n\rho(1-u)1_{d}$ and $w_{\PH}=1_{d}$. \\

Therefore, using \eqref{interm} and the matrix determinant lemma, we obtain as before
\begin{eqnarray}
	\frac{|\Sigma_{d}(\rho)|}{|\Sigma_{\PH}(u,\rho)|}
	&=& |\Sigma_{d}(\rho)\Sigma_{\PH}(u,\rho)^{-1}|\nonumber\\
	&=& |A_{\PH}|\big(1+w_{\PH}^{\T}A_{\PH}^{-1}v_{\PH}\big)\nonumber\\
	&=& \big[1+n(1-\rho)(1-u)\big]^{d} \times \big[1+\frac{n\rho(1-u)}{1+n(1-\rho)(1-u)}1_{d}^{\T}1_{d}\big]\nonumber\\
	&=& \big[1+n(1-\rho)(1-u)\big]^{d-1}\times \big[1+n\{1+(d-1)\rho\}(1-u)\big].\label{PHB_RISK_3}
\end{eqnarray}
Combining (\ref{PHB_RISK_2}) and (\ref{PHB_RISK_3}), the rest of the proof follows immediately. This completes the proof of Proposition \ref{PROP_INTRISK_COMPND_SYMM}.

\section*{Proof of Theorem \ref{THM_RISK_COMP_COMPND_SYMM}:}
\noindent{\textit{Proof:}} By definition, the risk difference function $H(\rho)$ is given by
\small \begin{eqnarray}
	H(\rho)
	&=& R(\hat{\mu}_{\PHB},\rho)-R(\hat{\mu}_{\HB},\rho)\nonumber\\
	&=& \frac{d-3}{n\{1+n(1-\rho)\}}-
	\frac{(d-2)^2}{2n}\int_{0}^{1} \frac{u^{\frac{d-2}{2}-1}}{\surd{\xi(u;\rho)}}du,\label{RISK_DIFF}
\end{eqnarray}
\normalsize
for $-1/(d-1)<\rho<1$, where $\xi(u;\rho)$ is already defined in Theorem \ref{THM_RISK_COMP_COMPND_SYMM}. \\

Let us now rewrite $H(\rho)$ in (\ref{RISK_DIFF}) as
\begin{eqnarray}
	H(\rho)
	&=&  
	\frac{\big[(d-3)-(d-2)\int_{0}^{1}h_{\rho}(u)f_{\rho}(u)du\big]}{n[1+n(1-\rho)]}
	\label{RISK_DIFF_1}
\end{eqnarray}
where
\begin{equation}
	h_{\rho}(u)=\bigg[\frac{1+n(1-\rho)(1-u)}{1+n(1+(d-1)\rho)(1-u)}\bigg]^{1/2}, \quad u\in[0,1]\label{DEFN_h}
\end{equation}
and
\begin{equation}
	f_{\rho}(u)=\frac{(d-2)[1+n(1-\rho)]u^{(d-2)/2-1}}{2\big[1+n(1-\rho)(1-u)\big]^{d/2}}, \quad u\in[0,1].\label{DEFN_f}
\end{equation}
We observe that for each fixed $\rho$, $f_{\rho}(u)\geq 0$ for all $u\in[0,1]$ and $\int_{0}^{1}f_{\rho}(u)du=1$ (using (\ref{RISK_HB_6})). Using this observation and (\ref{RISK_DIFF_1}), let us rewrite $H(\rho)$ as
\begin{eqnarray}
	H(\rho)
	&=&\frac{d-2}{n[1+n(1-\rho)]}\int_{0}^{1}(\surd{\alpha_{d}}-h_{\rho}(u))f_{\rho}(u)du,\nonumber 
\end{eqnarray}
where $\alpha_{d}=\{(d-3)/(d-2)\}^2$.

Note that $H(\rho)<0$ if $h_{\rho}(u)>\surd{\alpha_{d}}$ for all $u\in(0,1)$, that is, if and only if, $ \mbox{ for each } u\in(0,1)$
\begin{equation}
	1+n(1-u)\bigg[1-\rho\frac{1+(d-1)\alpha_{d}}{1-\alpha_{d}}\bigg]> 0.\label{RISK_DIFF_2}
\end{equation}
Set, $\rho_{L}=(1-\alpha_{d})(1+n^{-1})/\{1+(d-1)\alpha_{d}\}\in(0,1)$, provided $2d<nd(d-3)^2+5$. One may easily verify that the preceding condition would automatically satisfy for every $d\ge 5$ and any $n\geq 1$. Observe that, when $\rho\leq\rho_{L}$, the quantity
\begin{equation}
	1+n(1-u)\bigg[1-\rho\frac{1+(d-1)\alpha_{d}}{1-\alpha_{d}}\bigg] \nonumber
\end{equation}
is 
\begin{eqnarray}
	&\geq& 1+n(1-u)\bigg[1-\rho_{L}\frac{1+(d-1)\alpha_{d}}{1-\alpha_{d}}\bigg]\nonumber\\
	&=& u\nonumber\\
	&>& 0,\nonumber
\end{eqnarray}
for all $u \in (0,1)$. Thus (\ref{RISK_DIFF_2}) will be satisfied by every $\rho\leq\rho_{L}$. Hence $H(\rho)<0$ for all $-1/(d-1)<\rho\leq\rho_{L}$.

Let us now consider the case $\rho>0$. Define,
\begin{equation}\label{DEFN_I}
	I(\rho)=\int_{0}^{1}h_{\rho}(u)f_{\rho}(u)du, \ \rho\in(0,1),
\end{equation}
so that
\begin{eqnarray}
	H(\rho)
	&=&\frac{1}{n[1+n(1-\rho)]}\big[(d-3)-(d-2)I(\rho)].\label{RISK_DIFF_3}
\end{eqnarray}
We claim that, for all $d\ge 5$ and for each $ n\ge 1$, $\lim_{\rho\uparrow 1}H(\rho)>0$ which is equivalent to saying that $\lim_{\rho\uparrow 1}I(\rho)< \surd{\alpha_{d}}$, where
\begin{equation}\label{limit_I_rho}
	\lim\limits_{\rho\uparrow 1} I(\rho)=\int_{0}^{1} \frac{(d-2)u^{\frac{d-2}{2}-1}}{2\sqrt{1+nd(1-u)}}du.
\end{equation}
The right-hand side of (\ref{limit_I_rho}) follows by using (\ref{DEFN_h}), (\ref{DEFN_f}) and (\ref{DEFN_f}), together with an application of the dominated convergence theorem. Since
\begin{equation}
	\int_{0}^{1} \frac{(d-2)u^{\frac{d-2}{2}-1}}{2\sqrt{1+nd(1-u)}}du \le \int_{0}^{1} \frac{(d-2)u^{\frac{d-2}{2}-1}}{2\sqrt{1+d(1-u)}}du,\nonumber
\end{equation}
for all $n\ge1$, it would be enough to show that
\begin{equation}\label{Contrapositive_Irho_alpha}
	\int_{0}^{1} \frac{(d-2)u^{\frac{d-2}{2}-1}}{2\sqrt{1+d(1-u)}}du < \surd{\alpha_{d}}
\end{equation}
for all $d\ge5$ to establish our aforesaid claim. Towards that, let us first write
\begin{equation}
	J(d)=\int_{0}^{1} \frac{(d-2)u^{\frac{d-2}{2}-1}}{2\sqrt{1+d(1-u)}}du.\label{Jd_defn}
\end{equation}
Now we rewrite the integral $J(d)$ in (\ref{Jd_defn}) as
\begin{eqnarray}
	J(d)
	&=& \int_{0}^{1} \frac{(d-2)(1-u)^{\frac{d-2}{2}-1}}{2\sqrt{1+ud}}du.\label{Jd_defn_equiv_1}
\end{eqnarray}
Using (\ref{Jd_defn_equiv_1}), $J(d)$ can be bounded above as
\begin{eqnarray}
	J(d)
	&=&\bigg(\int_{0}^{\beta}+\int_{\beta}^{1}\bigg)\frac{(d-2)(1-u)^{\frac{d-2}{2}-1}}{2\sqrt{1+ud}}du\nonumber\\
	&<& \int_{0}^{\beta}\frac{(d-2)}{2}(1-u)^{\frac{d-2}{2}-1}du+\int_{\beta}^{1}\frac{(d-2)(1-u)^{\frac{d-2}{2}-1}}{2\sqrt{1+\beta d}}du\nonumber\\
	&=& 1- (1-\beta)^{\frac{d-2}{2}}\bigg[1-\frac{1}{\sqrt{1+d\beta}}\bigg],\label{Jd_UB1}
\end{eqnarray}
for any $0<\beta<1$. Since $\beta\in(0,1)$ is arbitrary, let us take $\beta=d^{-1}$ in (\ref{Jd_UB1}) so that
\begin{eqnarray}
	J(d)
	&<& 1- \bigg(1-\frac{1}{\surd{2}}\bigg)\bigg(1-\frac{1}{d}\bigg)^{\frac{d}{2}-1}\nonumber\\
	&<& 1- \bigg(1-\frac{1}{\surd{2}}\bigg)\bigg(1-\frac{1}{d}\bigg)^{\frac{d}{2}}.\label{Jd_UB2}
\end{eqnarray}
The last step in (\ref{Jd_UB2}) follows from the fact $\big(1-1/d\big)^{-1}>1$. Since the function $d\mapsto \big(1-1/d\big)^{d}$ with $d\ge 5$, is strictly increasing in $d$, we obtain, for all $d\ge 5$
\begin{eqnarray}
	J(d)
	&<& 1- \bigg(1-\frac{1}{\surd{2}}\bigg)\bigg(\frac{4}{5}\bigg)^{\frac{5}{2}}\nonumber\\
	&\approx& 0.8323381. \label{Jd_UB3}
\end{eqnarray}
Since $\surd{\alpha_{d}}=(d-3)/(d-2)$ is strictly increasing in $d$ with $\surd{\alpha_{8}} \approx 0.8333333$, using (\ref{Jd_UB3}), it follows that $J(d)<\surd{\alpha_{d}}$ for all $d\ge8$. Again, by using the transformation $v=ud/(d+1)$ on the right hand side of (\ref{Jd_defn}), let us rewrite the integral $J(d)$ as
\begin{eqnarray}
	J(d)
	&=& \frac{d-2}{2\sqrt{d+1}}\bigg(1+\frac{1}{d}\bigg)^{\frac{d}{2}-1}\int_{0}^{\frac{d}{d+1}}\frac{v^\frac{d-2}{2}-1}{\sqrt{1-v}}dv.\label{Jd_defn_equiv_2}
\end{eqnarray}
Using properties of incomplete beta functions, it can be verified easily from (\ref{Jd_defn_equiv_2}) that $J(d) < \surd{\alpha_{d}}$ when $d=5,6 \mbox{ and } 7$. Combining the preceding observations, we see that (\ref{Contrapositive_Irho_alpha}) is satisfied by all $d\ge5$. This establishes our aforesaid claim.\par
Applying the dominated convergence theorem, it follows that the function $I(\rho)$ and hence $H(\rho)$ are continuous in $\rho$. Hence, $H(\rho)$ must be positive for all values of $\rho$ sufficiently close to $1$. Therefore, applying the mean value theorem for continuous functions, it follows that there exists some $\rho^{\ast}\in(\rho_{L},1)$ such that $H(\rho^{\ast})=0$.\par
We next claim that the function $H(\rho)$ is strictly increasing in $\rho$ over $(0,1)$ so that $H(\rho)$ changes it sign exactly once at $\rho=\rho^{\ast}$. Toward that end, we first observe that the function $\rho\mapsto\frac{1}{1+n(1-\rho)}$ is strictly increasing in $\rho$ over $(0,1)$. Therefore, in view of (\ref{RISK_DIFF_3}), it would suffice to show that $I(\rho)$ strictly decreases in $\rho\in(0,1)$. Now one can rewrite $I(\rho)$ as
\begin{equation}\label{I_rho_alt}
	I(\rho)=1-\int_{0}^{1}\bigg[\frac{u}{1+n(1-\rho)(1-u)}\bigg]^{\frac{d-2}{2}}\frac{\partial}{\partial u} h_{\rho}(u)du.
\end{equation}
Therefore, differentiating both sides of (\ref{I_rho_alt}) with respect to $\rho$ and applying Leibnitz's theorem, we obtain after some algebraic manipulations
\begin{eqnarray}\label{I_prime}
	\dot{I}(\rho)&=&-\int_{0}^{1} \bigg\{\frac{u}{1+n(1-\rho)(1-u)}\bigg\}^{\frac{d}{2}-2}\times \bigg[\bigg(\frac{d}{2}-1\bigg)n(1-u)\frac{\partial}{\partial u} h_{\rho}(u)\ +\nonumber\\
	&& \qquad \ \frac{u}{1+n(1-\rho)(1-u)}\frac{\partial^2}{\partial \rho \partial u} h_{\rho}(u)du \bigg].
\end{eqnarray}
In (\ref{I_prime}) above, $\dot{I}(\rho)$ stands for the first-order derivative of $I(\rho)$ with respect to $\rho$. Now differentiating the log--transformation of $h_{\rho}(u)$ with respect to $u$, we obtain for each fixed $\rho\in(0,1)$
\begin{equation}\label{PART_DERI_h_u}
	\frac{\partial}{\partial u} h_{\rho}(u)
	=\frac{h_{\rho}(u)}{2(1-u)}\bigg[\frac{1}{1+n(1-\rho)(1-u)} - \frac{1}{1+n(1+(d-1)\rho)(1-u)}\bigg],
\end{equation}
which is strictly positive for all $u\in(0,1)$. Using (\ref{PART_DERI_h_u}), for each fixed $u\in(0,1)$, the functions $(d/2-1)n(1-u)\partial h_{\rho}(u)/\partial u$ and $u/\{1+n(1-\rho)(1-u)\} $ are strictly positive for all $\rho\in(0,1)$. Therefore, in view of (\ref{I_prime}) and the preceding observations, it would be enough to establish,
for each fixed $0<u<1$,
\begin{equation}\label{h_2nd_derivative_criteria}
	\frac{\partial^2}{\partial \rho \partial u} h_{\rho}(u) > 0
\end{equation}
for all values of $\rho\in(0,1)$ to establish the aforesaid claim. This is so because if (\ref{h_2nd_derivative_criteria}) were true, the integrand on the right-hand side of (\ref{I_prime}) is strictly positive for each fixed $u\in(0,1)$. Hence, $\dot{I}(\rho)$ is strictly negative for all $\rho\in(0,1)$, from which the aforesaid claim follows. So, in order to show (\ref{h_2nd_derivative_criteria}), let us first rewrite (\ref{PART_DERI_h_u}) as
\begin{equation}\label{PART_DERI_h_u_equiv}
	\frac{\partial}{\partial u} h_{\rho}(u)
	=\frac{h_{\rho}(u)\phi_{\rho}(u)}{2(1-u)}
\end{equation}
where for each fixed $u\in (0,1)$ and each fixed $\rho\in(0,1)$
\begin{equation}\label{phi_defn}
	\phi_{\rho}(u)=\frac{1}{1+n(1-\rho)(1-u)} - \frac{1}{1+n(1+(d-1)\rho)(1-u)}.
\end{equation}
Now for each fixed $u\in(0,1)$
\begin{eqnarray}
	\frac{\partial}{\partial \rho} h_{\rho}(u)
	&=&-\frac{h_{\rho}(u)}{2}\bigg[\frac{n(1-u)}{1+n(1-\rho)(1-u)} + \nonumber \\
	&& \frac{n(d-1)(1-u)}{1+n(1+(d-1)\rho)(1-u)}\bigg].\label{PART_DERI_h}\
\end{eqnarray}
Again differentiating both sides of (\ref{phi_defn}) with respect to $\rho$ we obtain for each $u\in(0,1)$
\begin{eqnarray}
	\frac{\partial}{\partial \rho} \phi_{\rho}(u)
	&=& \frac{n(1-u)}{[1+n(1-\rho)(1-u)]^2}+\frac{n(d-1)(1-u)}{[1+n(1+(d-1)\rho)(1-u)]^2}.
\end{eqnarray}
Next differentiating both sides of (\ref{PART_DERI_h_u_equiv}) with respect to $\rho$, we get
\begin{eqnarray}
	\frac{\partial^2}{\partial \rho \partial u} h_{\rho}(u)
	&=&\frac{1}{2(1-u)}\bigg[\phi_{\rho}(u)\frac{\partial}{\partial \rho} h_{\rho}(u) + h_{\rho}(u)\frac{\partial}{\partial \rho}\phi_{\rho}(u) \bigg]\nonumber\\
	&=& \frac{nh_{\rho}(u)}{2}\bigg[-\frac{1}{2}\bigg\{\frac{1}{1+n(1-\rho)(1-u)} - \frac{1}{1+n(1+(d-1)\rho)(1-u)}\bigg\}\ \times \nonumber \\
	&& \qquad \bigg\{\frac{1}{1+n(1-\rho)(1-u)} + \frac{d-1}{1+n(1+(d-1)\rho)(1-u)}\bigg\} \ +\nonumber\\
	&& \qquad \bigg\{\frac{1}{\{1+n(1-\rho)(1-u)\}^2}+\frac{d-1}{\{1+n(1+(d-1)\rho)(1-u)\}^2}\bigg\} \bigg] \nonumber\\
	&=& \frac{nh_{\rho}(u)}{4}\bigg[\frac{1}{\{1+n(1-\rho)(1-u)\}^2} + \frac{3(d-1)}{\{1+n(1+(d-1)\rho)(1-u)\}^2}\bigg].\nonumber
\end{eqnarray}
which clearly satisfies (\ref{h_2nd_derivative_criteria}) for each fixed  $0<u<1$ and for each fixed $\rho\in(0,1)$. This establishes the fact that $I(\rho)$ strictly decreases as $\rho$ increases over $(0,1)$ whence our last claim follows immediately. Therefore, $H(\rho^{\ast})=0$ for some unique $\rho^{\ast}\in(\rho_{L},1)$.\par
Our next aim would be to obtain a non-trivial $\rho_{U}\in(\rho_{L},1)$ such that $H(\rho_{U})>0$ which ensures that the unique $\rho^{\ast}\in(\rho_{L},\rho_{U})$.  When $\rho >0$ is held fixed, $h_{\rho}(u)$ is continuous in $u$ over $[0,1]$. Again, from (\ref{PART_DERI_h_u}), it follows $\partial h_{\rho}(u)/\partial u$ is strictly positive for all $u\in(0,1)$, when $\rho\in(0,1)$ is fixed. Therefore, for each fixed $\rho >0$, the function $h_{\rho}(u)$ strictly increases as $u$ increases over $[0,1]$.
Using these observations and applying the generalized mean value theorem for Riemann integrals, we obtain
\begin{eqnarray}
	I(\rho)
	&=& h_{\rho}(u_{0}(\rho))\int_{0}^{1}f_{\rho}(u)du\nonumber\\
	&=& h_{\rho}(u_{0}(\rho)),\label{RISK_DIFF_4}
\end{eqnarray}
for some $u_{0}(\rho)\in(0,1)$, that depends on the corresponding $\rho$. Here the strict monotonicity property of $h_{\rho}(u)$ as a function of $u$ ensures $u_{0}(\rho)\in(0,1)$.This, in conjunction with (\ref{RISK_DIFF_4}) and the definition of the function $h_{\rho}$, implies that $I(\rho)$ strictly lies in between $0$ and $1$. In a short while, we shall see the aforesaid observation is crucial for completing the rest of the arguments that follow.

Next, we claim $u_{0}(\rho)$'s are uniformly bounded away from $1$, that is, there exists some $\delta_{0}\in(0,1)$, independent of any $\rho\in(0,1)$ and possibly depending on $d$  and $n$, such that $u_{0}(\rho) <1-\delta_{0}$ for all $\rho\in(0,1)$. This can be proven based on a contra-positive argument as follows. If possible, let us assume our claim is not true. Then there does not exist any $\delta\in(0,1)$ such that $u_{0}(\rho)\leq 1-\delta$ holds for all $\rho\in(0,1)$. This means, for every $\delta\in(0,1)$, there exists some $\rho_{\delta}\in(0,1)$ such that $u_{0}(\rho_{\delta})> 1-\delta$ whence
\begin{eqnarray}
	I(\rho_{\delta})
	&=& h_{\rho_{\delta}}(u_{0}(\rho_{\delta}))\nonumber\\
	&>& h_{\rho_{\delta}}(1-\delta)\nonumber\\
	&=& \bigg[\frac{1+n(1-\rho_{\delta})\delta}{1+n(1+(d-1)\rho_{\delta})\delta}\bigg]^{1/2}.\label{RISK_DIFF_5}
\end{eqnarray}
The second step in the above chain of inequalities (\ref{RISK_DIFF_5}) follows from the non--decreasing property of $h_{\rho_{\delta}}$. Now consider a sequence $\{\delta_{k}\}$ in $(0,1)$ such that $\delta_{k}\rightarrow0$ as $k\rightarrow\infty$. Then, using (\ref{RISK_DIFF_5}) we obtain
\begin{eqnarray}
	\liminf_{k\rightarrow\infty} I(\rho_{\delta_{k}})
	&\geq& \liminf_{k\rightarrow\infty} \bigg[\frac{1+n(1-\rho_{\delta_{k}})\delta_{k}}{1+n(1+(d-1)\rho_{\delta_{k}})\delta_{k}}\bigg]^{1/2}\nonumber\\
	&=& 1.\nonumber
\end{eqnarray}
This implies $I(\rho_{\delta_{k}})\geq 1$ for all sufficiently large $k$ which is a contradiction since $I(\rho_{\delta_{k}}) < 1$ for all $k$. This establishes the aforesaid claim.\par
Our next claim is that the function $\rho\mapsto u_{0}(\rho)$ steadily increases in $\rho$ over $(0,1)$. From (\ref{PART_DERI_h}), we see that $\partial h{\rho}(u)/\partial\rho$ is strictly negative for all $u\in(0,1)$. Hence the function $\rho\mapsto h_{\rho}(u)$, $\rho\in(0,1)$, is strictly decreasing in $\rho$. Thus, inverse of the function $h_{\rho}$ exists so that $u_{0}(\rho)=h_{\rho}^{-1}(I(\rho))$. Moreover, the function $\rho\mapsto h^{-1}_{\rho}$ strictly decreases as $\rho$ increases in $(0,1)$. Also, the function $I(\rho)$ is strictly decreasing in $\rho$ increases over $(0,1)$. The aforesaid claim follows since the composition of two decreasing real-valued functions in one real variable is increasing.\par
Now, we observe
\begin{equation}\label{Condition_Positivity_1}
	H(\rho)>0 \mbox{ if and only if } I(\rho)< \surd{\alpha_{d}},
\end{equation}
that is, if and only if
\begin{equation}\label{Condition_Positivity_2}
	\rho > \frac{1-\alpha_{d}}{1+(d-1)\alpha_{d}}\big[1+\frac{1}{n(1-u_{0}(\rho))}\big].
\end{equation}

Define, $\delta_{\ast}=\inf_{\rho\in(0,1)}[1-u_{0}(\rho)]$. Since $u_{0}(\rho)$'s lie in $(0,1)$ and are uniformly bounded away from $1$ with respect to $\rho\in(0,1)$, $\delta_{\ast}$ must lie in $(0,1)$. Since $u_{0}(\rho)=h_{\rho}^{-1}(I(\rho))$ strictly increases in $\rho$ over $(0,1)$, we have
\begin{eqnarray}
	\delta_{\ast}
	&=& 1-\sup_{\rho\in(0,1)}u_{0}(\rho)\nonumber\\
	&=& 1-\lim\limits_{\rho\uparrow 1}u_{0}(\rho).\label{delta_equiv}
\end{eqnarray}
Now using (\ref{DEFN_h}) and (\ref{RISK_DIFF_4}), it can be verified
\begin{eqnarray}\label{u0_rho}
	u_{0}(\rho)
	&=& h_{\rho}^{-1}(I(\rho))\nonumber\\
	&=& 1 - \frac{1}{n}\frac{1-I^{2}(\rho)}{\{1+(d-1)\rho\}I^{2}(\rho)-(1-\rho)}.
\end{eqnarray}
Therefore, using (\ref{delta_equiv}) and (\ref{u0_rho}) we obtain
\begin{equation}\label{delta_exact}
	\delta_{\ast}=\frac{1}{dn}\bigg[\frac{1}{\lim\limits_{\rho\uparrow 1} I^2(\rho)}-1 \bigg].
\end{equation}
Let us now define
\begin{equation}\label{DEFN_rho_U}
	\rho_{U}=\frac{1-\alpha_{d}}{1+(d-1)\alpha_{d}}\bigg(1+\frac{1}{n\delta_{\ast}}\bigg).
\end{equation}
We claim $\rho_{U}<1$. We shall prove this using a contra--positive argument. On the contrary, let us assume the claim to be false, that is, $\rho_{U}\ge1$, which would be true if and only if $\delta_{\ast}\le(\alpha_{d}^{-1}-1)/(nd)$. In view of (\ref{delta_exact}), the last assertion would be equivalent to saying $\lim_{\rho\uparrow 1} I(\rho)\ge\surd{\alpha_{d}}$ which contradicts (\ref{Contrapositive_Irho_alpha}) when $d\ge5$ and $n=1$. Hence $\rho_{U}$ must be smaller than $1$.\par
Finally, we observe
\begin{equation}
	\rho_{U} = \sup_{\rho\in(0,1)}\frac{1-\alpha_{d}}{1+(d-1)\alpha_{d}}\bigg[1+\frac{1}{n(1-u_{0}(\rho))}\bigg].\nonumber
\end{equation}
Clearly, every $\rho\geq\rho_{U}$ satisfies (\ref{Condition_Positivity_2}). Hence $H(\rho)>0$ for every such values of $\rho$. In particular, $H(\rho_{U})>0$. As argued before, this ensures the unique $\rho^{\ast}$ lies inside $(\rho_{L},\rho_{U})$, thus concluding the proof of Theorem \ref{THM_RISK_COMP_COMPND_SYMM}.

\section*{Proof of Proposition \ref{PROP_INTRISK_GEN}:}
\noindent{\textit{Proof:}} Arguing exactly the same way as in the proof of Proposition \ref{PROP_INTRISK_COMPND_SYMM}, one can show
\begin{eqnarray}\label{RISK_HB_GEN_1}
	R(\hat{\mu}_{\HB},G_{0})
	&=& \int_{\mathbb{R}^d} E_{Y\mid\mu}\lVert \hat{\mu}_{\HB}-\mu\rVert^2_{2}dG_{0}(\mu)\nonumber\\
	&=& \frac{d}{n}-\frac{(d-3)^2}{2n}\int_{0}^{1}u^{(d-3)/2-1} \bigg\{\frac{\mid\widetilde{\Sigma}_{\H}(u)\mid}{\mid\Sigma\mid}\bigg\}^{1/2}du,
\end{eqnarray}
where $\widetilde{\Sigma}_{\H}(u)=\big[n(1-u)(I_{d}-d^{-1}1_{d}1_{d}^{\T})+\Sigma^{-1}\big]^{-1}$ is a $d\times d$ positive definite matrix for each fixed $u\in[0,1]$.

Next, for each fixed $u\in[0,1]$
\begin{eqnarray}\label{RISK_HB_GEN_2}
	\Sigma\widetilde{\Sigma}_{\H}(u)^{-1}
	&=&n(1-u)\Sigma(I_{d}-d^{-1}1_{d}1_{d}^{\T})+I_{d}\nonumber\\
	&=& [n(1-u)\Sigma+I_{d}]+\big\{-\frac{n(1-u)}{d}\big\}\Sigma1_{d}1_{d}^{\T}\nonumber\\
	&=& \widetilde{A}_{\H}+\widetilde{v}_{\H}\widetilde{w}_{\H}^{\T},
\end{eqnarray}
where $\widetilde{A}_{\H}=n(1-u)\Sigma+I_{d}$, $\widetilde{v}_{\H}=-d^{-1}n(1-u)\Sigma1_{d}$ and $\widetilde{w}_{\H}=1_{d}$.

As before, using the matrix determinant lemma we obtain
\begin{eqnarray}\label{RISK_HB_GEN_3}
	\frac{\mid\Sigma\mid}{\mid\widetilde{\Sigma}_{\H}(u)\mid}
	&=& \mid\Sigma\widetilde{\Sigma}_{\H}(u)^{-1}\mid\nonumber\\
	&=& \mid\widetilde{A}_{\H}\mid\big(1+\widetilde{w}_{\H}^{\T}\widetilde{A}_{\H}^{-1}\widetilde{v}_{\H}\big)\nonumber\\
	&=& \xi_{1}(u),
\end{eqnarray}
where $\xi_{1}(u)=\mid\widetilde{A}_{\H}\mid\big(1+\widetilde{w}_{\H}^{\T}\widetilde{A}_{\H}^{-1}\widetilde{v}_{\H}\big)$, $u\in[0,1]$. Using the singular value decomposition of $\Sigma$, we may write $\Sigma=P\Lambda P^{\T}$, where $P$ and $\Lambda$ have already been defined in the statement of Proposition \ref{PROP_INTRISK_GEN}. Using this observation, and noting $P P^{\T}=I_{d}$, one may rewrite $\widetilde{A}_{\H}$ as
\begin{eqnarray}\label{RISK_HB_GEN_4}
	\widetilde{A}_{\H}
	&=& n(1-u)P\Lambda P^{\T}+P P^{\T}\nonumber\\
	&=& P\big[n(1-u)\Lambda+I_{d}\big]P^{\T}.
\end{eqnarray}
Using (\ref{RISK_HB_GEN_4}) we obtain
\begin{eqnarray}\label{RISK_HB_GEN_5}
	|\widetilde{A}_{\H}|
	&=& |P||n(1-u)\Lambda+I_{d}||P^{\T}|\nonumber\\
	&=& |PP^{\T}||n(1-u)\Lambda+I_{d}|\nonumber\\
	&=& |n(1-u)\Lambda+I_{d}|\nonumber\\
	&=& |diag(n(1-u)\lambda_{1}+1,\dots,n(1-u)\lambda_{d}+1)|\nonumber\\
	&=& \prod_{j=1}^{d}\big\{1+n(1-u)\lambda_{j}\big\},
\end{eqnarray}
where $diag(\beta_{1},\dots,\beta_{d})$ denotes a diagonal matrix with diagonal entries $\beta_{1},\dots,\beta_{d}$. Since $\Sigma=P\Lambda P^{\T}$, and $Z=P^{\T}1_{d}$ we get
\small \begin{eqnarray}\label{RISK_HB_GEN_6}
	1+\widetilde{w}_{\H}^{\T}\widetilde{A}_{\H}^{-1}\widetilde{v}_{\H}
	&=& 1-\frac{n(1-u)}{d}1_{d}^{\T}\bigg[n(1-u)\Sigma+I_{d}\bigg]^{-1}\Sigma1_{d}\nonumber\\
	&=& 1-\frac{n(1-u)}{d}1_{d}^{\T}P\bigg[n(1-u)\Lambda+I_{d}\bigg]^{-1}\Lambda P^{\T}1_{d}\nonumber\\
	&=& 1-\frac{n(1-u)}{d}Z^{\T}diag\bigg(\frac{\lambda_{1}}{1+n(1-u)\lambda_{1}},\nonumber \\ && \dots, \frac{\lambda_{d}}{1+n(1-u)\lambda_{d}}\bigg)Z\nonumber\\
	&=& 1-\frac{n(1-u)}{d}\sum_{j=1}^{d}\frac{\lambda_{j}Z_{j}^2}{1+n(1-u)\lambda_{j}}.
\end{eqnarray}
\normalsize 
Combining (\ref{RISK_HB_GEN_1}), (\ref{RISK_HB_GEN_3}), (\ref{RISK_HB_GEN_5}) and (\ref{RISK_HB_GEN_6}), the risk expression in (\ref{RISK_HB_GENERAL_CASE}) becomes immediate. An exactly similar approach leads to the integrated risk profile of $\hat{\mu}_{\PHB}$ in (\ref{RISK_PHB_GENERAL_CASE}). This completes the proof of Proposition \ref{PROP_INTRISK_GEN}.

\section*{Proof of Theorem \ref{THM_RISK_COMPARISON_GENERAL}:}
\noindent{\textit{Proof:}} From Proposition \ref{PROP_INTRISK_GEN}, the integrated risk function of $\hat{\mu}_{\HB}$ is given by
\begin{equation}
	R(\hat{\mu}_{\HB},G_{0})=\frac{d}{n}-\frac{(d-3)^2}{2n}\int_{0}^{1}\frac{u^{(d-3)/2-1}}{\surd{\xi_{1}(u)}}du,\label{GEN_RISK_HB}
\end{equation}
where for $u\in[0,1]$ we have
\begin{eqnarray}\label{GEN_XI_1}
	\xi_{1}(u)
	&=&\bigg[1-\frac{n(1-u)}{d}\sum_{j=1}^{d}\frac{\lambda_{j}Z_{j}^2}{1+n(1-u)\lambda_{j}} \bigg]\prod_{j=1}^{d}\big\{1+n(1-u)\lambda_{j}\big\},
\end{eqnarray}
with $\lambda_{j}=\lambda_{0,j}$ for $j=1,\dots,d-1$, and $\lambda_{d}=\lambda_{0,d}+\nu$. 
Since $P=P_{0}$, $Z=P^{\T}1_{d}=P_{0}^{\T}1_{d}$. Again, since $\nu>0$ we have
\begin{equation}
	\frac{n(1-u)}{d}\frac{(\lambda_{0,d}+\nu)Z_{d}^2}{1+n(1-u)(\lambda_{0,d}+\nu)} > \frac{n(1-u)}{d}\frac{\lambda_{0,d}Z_{d}^2}{1+n(1-u)\lambda_{0,d}},\label{GEN_XI_1_UB1}
\end{equation}
for all $u\in(0,1)$. Using (\ref{GEN_XI_1}) and (\ref{GEN_XI_1_UB1}), and the preceding observations we obtain
\begin{eqnarray}\label{GEN_XI_1_UB2}
	\xi_{1}(u) &<& \bigg[1-\frac{n(1-u)}{d}\sum_{j=1}^{d}\frac{\lambda_{0,j}Z_{j}^2}{1+n(1-u)\lambda_{0,j}} \bigg]\prod_{j=1}^{d}\big\{1+n(1-u)\lambda_{j}\big\}\nonumber\\
	&=& \xi_{0,1}(u)\times \frac{1+n(1-u)(\lambda_{0,d}+\nu)}{1+n(1-u)\lambda_{0,d}},
\end{eqnarray}
where
\begin{equation}
	\xi_{0,1}(u)=
	\bigg[1-\frac{n(1-u)}{d}\sum_{j=1}^{d}\frac{\lambda_{0,j}Z_{j}^2}{1+n(1-u)\lambda_{0,j}} \bigg]\prod_{j=1}^{d}\big\{1+n(1-u)\lambda_{j}\big\}, \ u \in (0,1).\label{XI0_1_defn}
\end{equation}
Combining (\ref{GEN_RISK_HB}) and (\ref{GEN_XI_1_UB2}), we get
\begin{eqnarray}\label{GEN_RISK_HB_UB1}
	R(\hat{\mu}_{\HB},G_{0})
	&<& \frac{d}{n}-\frac{(d-3)^2}{2n}\int_{0}^{1}\frac{u^{(d-3)/2-1}}{\surd{\xi_{0,1}(u)}}\times \bigg[\frac{1+n(1-u)\lambda_{0,d}}{1+n(1-u)(\lambda_{0,d}+\nu)}\bigg]^{\frac{1}{2}} du\nonumber\\
	&=& \frac{d}{n}-\frac{(d-3)^2}{2n}\int_{0}^{1}\frac{u^{(d-3)/2-1}}{\surd{\xi_{0,1}(u)}}\times \psi_{\rho}(u)du,
\end{eqnarray}
where 
\begin{equation}
	\psi_{\rho}(u)=\bigg[\frac{1+n(1-u)\lambda_{0,d}}{1+n(1-u)(\lambda_{0,d}+\nu)}\bigg]^{\frac{1}{2}}, \ u \in (0,1).\nonumber
\end{equation}
Observe that, for each fixed $\rho\in(-1/(d-1),1)$
\begin{equation}
	\psi_{\rho}(u)
	=\bigg[\frac{1+n(1-u)\{1+(d-1)\rho\}}{1+n(1-u)\{1+(d-1)\rho+\nu\}}\bigg]^{\frac{1}{2}}, \ u \in (0,1).\nonumber
\end{equation}
Therefore, when $\rho\in(-1/(d-1),1)$ is kept fixed, taking the log--transformation of $\psi_{\rho}(u)$ and differentiating both sides with respect to $u$ we obtain
\begin{eqnarray}
	\frac{\partial}{\partial u} \psi_{\rho}(u)
	&=& \frac{\psi_{\rho}(u)}{2} \bigg[\frac{-n\{1+(d-1)\rho\}}{1+n(1-u)\{1+(d-1)\rho\}}+\frac{n\{1+(d-1)\rho+\nu\}}{1+n(1-u)\{1+(d-1)\rho+\nu\}}\bigg]\notag\\
	&=&\frac{\psi_{\rho}(u)}{2(1-u)}\bigg[\frac{1}{1+n(1-u)\{1+(d-1)\rho\}}-\frac{1}{1+n(1-u)\{1+(d-1)\rho+\nu\}}\bigg].\notag
\end{eqnarray}
Since $-1/(d-1)<\rho<1$, $\partial\psi_{\rho}(u)/\partial u  $ is strictly positive for each $u\in(0,1)$. Thus, when $\rho\in(-1/(d-1),1)$ is held fixed, $\psi_{\rho}(u)$ is strictly increasing in $u$ over $(0,1)$. Also, given $\rho$, $\psi_{\rho}(u)$ is continuous in $u\in(0,1)$. Hence, $\inf_{u\in(0,1)}\psi_{\rho}(u)=\lim_{u\rightarrow0+}\psi_{\rho}(u)$. Using these observations together with (\ref{GEN_RISK_HB_UB1}), we have
\begin{eqnarray}
	R(\hat{\mu}_{\HB},G_{0})
	&<& \frac{d}{n}-\inf_{u\in(0,1)}\psi_{\rho}(u)\frac{(d-3)^2}{2n}\int_{0}^{1}\frac{u^{(d-3)/2-1}}{\surd{\xi_{0,1}(u)}}du\notag\\
	&=& \frac{d}{n}-\zeta_{1}(\rho)\frac{(d-3)^2}{2n}\int_{0}^{1}\frac{u^{(d-3)/2-1}}{\surd{\xi_{0,1}(u)}}du,\label{GEN_RISK_HB_UB2}
\end{eqnarray}
where $\zeta_{1}(\rho)=\inf_{u\in(0,1)}\psi_{\rho}(u)$. Observe that
\begin{equation}\label{ZETA_fn}
	\zeta_{1}(\rho)
	= \bigg[\frac{1+n\{1+(d-1)\rho\}}{1+n\{1+(d-1)\rho+\nu\}}\bigg]^{\frac{1}{2}}.
\end{equation}
Employing the arguments used in the proofs of Proposition \ref{PROP_INTRISK_COMPND_SYMM} and Proposition \ref{PROP_INTRISK_GEN} we obtain
\begin{equation}
	\xi_{0,1}(u)=\big[1+n(1-\rho)(1-u)\big]^{d-1},  \ u \in (0,1).\notag
\end{equation}
Using (\ref{RISK_HB_6}) we get
\begin{equation}\label{INTEGRAL_IDENTITY}
	\frac{(d-3)}{2}\int_{0}^{1}\frac{u^{(d-3)/2-1}}{\surd{\xi_{0,1}(u)}}du=\frac{1}{1+n(1-\rho)}.
\end{equation}
On combining (\ref{GEN_RISK_HB_UB2}), (\ref{ZETA_fn}) and (\ref{INTEGRAL_IDENTITY}), we obtain
\begin{equation}
	R(\hat{\mu}_{\HB},G_{0}) < \frac{d}{n}-\frac{d-3}{n[1+n(1-\rho)]}\zeta_{1}(\rho).\label{GEN_RISK_HB_UB3}
\end{equation}
Now the function $\rho\mapsto\zeta_{1}(\rho)$ is strictly increasing in $\rho$. Hence
\begin{eqnarray}
	R(\hat{\mu}_{\HB},G_{0})
	&<& \frac{d}{n}-\frac{d-3}{n[1+n(1-\rho)]}\inf_{\rho\in(-1/(d-1),1)}\zeta_{1}(\rho)\nonumber\\
	&=& \frac{d}{n}-\beta_{\ast}\frac{d-3}{n[1+n(1-\rho)]},\label{GEN_RISK_HB_UB4}
\end{eqnarray}
where $\beta_{\ast}=\inf_{\rho\in(-1/(d-1),1)}\zeta_{1}(\rho)=(1+n\nu)^{-1/2}.$\\

Again, from Proposition \ref{PROP_INTRISK_GEN}, the risk function of $\hat{\mu}_{\PHB}$ is given by

\begin{eqnarray}
	R(\hat{\mu}_{\PHB},G_{0})
	&=&\frac{d}{n}-\frac{(d-2)^2}{2n}\int_{0}^{1}\frac{u^{(d-2)/2-1}}{\prod_{j=1}^{d}\big\{1+n(1-u)\lambda_{j}\big\}^{1/2}}du\nonumber\\
	&>& \frac{d}{n}-\frac{(d-2)^2}{2n}\int_{0}^{1}\frac{u^{(d-2)/2-1}}{\prod_{j=1}^{d}\big\{1+n(1-u)\lambda_{0,j}\big\}^{1/2}}du\nonumber\\
	&=& \frac{d}{n}-\frac{(d-2)^2}{2n}\int_{0}^{1} \frac{u^{\frac{d-2}{2}-1}}{\surd{\xi(u;\rho)}}du,\label{GEN_RISK_PHB_LB1}
\end{eqnarray}
\normalsize
where $\xi(u;\rho)$ is already defined in Theorem \ref{THM_RISK_COMP_COMPND_SYMM}.\\

Subtracting (\ref{GEN_RISK_PHB_LB1}) from (\ref{GEN_RISK_HB_UB4}) we obtain 
\begin{eqnarray}
	R(\hat{\mu}_{\HB},G_{0})- R(\hat{\mu}_{\PHB},G_{0})
	&<& \frac{(d-2)^2}{2n}\int_{0}^{1} \frac{u^{\frac{d-2}{2}-1}}{\surd{\xi(u;\rho)}}du-\beta_{\ast}\frac{d-3}{n[1+n(1-\rho)]}\label{GEN_RISK_DIFF_UB}
\end{eqnarray}
\normalsize
for $-1/(d-1)<\rho<1$.\\

Writing
\begin{equation}
	K(\rho)=\frac{(d-2)^2}{2n}\int_{0}^{1} \frac{u^{\frac{d-2}{2}-1}}{\surd{\xi(u;\rho)}}du-\beta_{\ast}\frac{d-3}{n[1+n(1-\rho)]}
\end{equation}
\normalsize 
for $-1/(d-1)<\rho<1$, we observe that $K(\rho)$ can be equivalently expressed as
\begin{eqnarray}
	K(\rho)
	&=&  
	\frac{\big[(d-2)\int_{0}^{1}h_{\rho}(u)f_{\rho}(u)du-(d-3)\beta_{\ast}\big]}{n[1+n(1-\rho)]}
\end{eqnarray}
where $h_{\rho}(u)$ and $f_{\rho}(u)$ have been defined already in (\ref{DEFN_h}) and (\ref{DEFN_f}), respectively.\par

Define $\widetilde{\rho}_{U}=(1-\alpha^{\ast}_{d})(1+n^{-1}\delta_{\ast}^{-1})/\{1+(d-1)\alpha^{\ast}_{d}\}$ and $\widetilde{\rho}_{L}=(1-\alpha^{\ast}_{d})(1+n^{-1})/\{1+(d-1)\alpha^{\ast}_{d}\}$, with $\alpha^{\ast}_{d}=\beta_{\ast}^2\alpha_{d}$ where the constants $\alpha_{d}$ and $\delta_{\ast}$ have already been defined in Theorem \ref{THM_RISK_COMP_COMPND_SYMM}. Clearly, both $\widetilde{\rho}_{U}$ and $\widetilde{\rho}_{L}$ are positive. Observe that $\widetilde{\rho}_{U}<1$ if and only if $\alpha^{\ast}_{d}(1+nd\delta_{\ast})>1$ which is equivalent to $\beta_{\ast}^{-2}<\alpha_{d}(1+nd\delta_{\ast})$. Since $\beta_{\ast}^{-2}=1+n\nu$, $\widetilde{\rho}_{U}<1$ if and only if $1+n\nu<\alpha_{d}(1+nd\delta_{\ast})$ which would be vacuously true since $0< \nu< n^{-1}\{\alpha_{d}(1+nd\delta_{\ast})-1\}$. The arguments used for establishing $\rho_{U}<1$ in the proof of Theorem \ref{THM_RISK_COMP_COMPND_SYMM} ensures that $\alpha_{d}(1+nd\delta_{\ast})-1>0$ for all $d\ge5$ and every $n\ge1$. Thus we have $\widetilde{\rho}_{U}<1$ whenever $d\ge5$ and $n\ge1$. Since $0<\delta_{\ast}<1$, we have $\widetilde{\rho}_{L}<\widetilde{\rho}_{U}$ whence $\widetilde{\rho}_{L}<1$ for all $d\ge5$ and $n\ge1$.\\

Using the preceding observations and employing precisely the same line of arguments used in the proof of Theorem \ref{THM_RISK_COMP_COMPND_SYMM} for analyzing the properties of $H(\rho)$, it follows that there exists a unique $\rho^{\ast}\in(\widetilde{\rho}_{L},\widetilde{\rho}_{U})$ such that
\begin{equation*}\label{RISK_DIFF_CONDITION_GEN_UB}
	\mathop{K(\rho)} \left\{
	\begin{array}{ll}
		> 0 & \mbox{ if}\; -\frac{1}{d-1}< \rho < \rho^{\ast},\\ \\
		= 0 & \mbox{ if}\; \rho = \rho^{\ast},\\ \\
		< 0 & \mbox{ if}\; \rho^{\ast} <\rho <1,
	\end{array}\right.
\end{equation*}
Since $R(\hat{\mu}_{\HB},G_{0})- R(\hat{\mu}_{\PHB},G_{0})< K(\rho)$, the proof of Theorem \ref{THM_RISK_COMPARISON_GENERAL} follows immediately.
%

\begin{thebibliography}{59}
\providecommand{\natexlab}[1]{#1}
\providecommand{\url}[1]{\texttt{#1}}
\expandafter\ifx\csname urlstyle\endcsname\relax
  \providecommand{\doi}[1]{doi: #1}\else
  \providecommand{\doi}{doi: \begingroup \urlstyle{rm}\Url}\fi

\bibitem[Allen et~al.(2018)Allen, Shin, Shelhamer, and
  Tenenbaum]{Allen_etal_2018}
Kevin~R. Allen, Hanzhang Shin, Evan Shelhamer, and Joshua~B. Tenenbaum.
\newblock Variadic meta-learning by bayesian nonparametric deep embedding.
\newblock In \emph{Advances in Neural Information Processing Systems
  (NeurIPS)}, Montr{\'e}al, Canada, 2018.

\bibitem[Amit and Meir(2018)]{Amit_Meir_2018}
Ron Amit and Ron Meir.
\newblock Meta-learning by adjusting priors based on extended pac-bayes theory.
\newblock In \emph{Proceedings of the 35th International Conference on Machine
  Learning (ICML)}, volume~80, Stockholm, Sweden, 2018. PMLR.

\bibitem[Berger(2013)]{Berger_2013}
James~O. Berger.
\newblock \emph{Statistical Decision Theory and Bayesian Analysis}.
\newblock Springer Science and Business Media, 2 edition, 2013.

\bibitem[Bernardo and Smith(2009)]{Bernardo_Smith_2009}
Jos{\'e}~M. Bernardo and Adrian F.~M. Smith.
\newblock \emph{Bayesian Theory}.
\newblock John Wiley \& Sons, 2009.

\bibitem[Bock(1975)]{Bock_1975}
M.~E. Bock.
\newblock Minimax estimators of the mean of a multivariate normal distribution.
\newblock \emph{Annals of Statistics}, 3:\penalty0 209--218, 1975.

\bibitem[Chen et~al.(2011)Chen, Polatkan, Sapiro, Dunson, and
  Carin]{Chen_etal_2011}
Bo~Chen, G.~Polatkan, G.~Sapiro, David~B. Dunson, and Lawrence Carin.
\newblock The hierarchical beta process for convolutional factor analysis and
  deep learning.
\newblock In \emph{Proceedings of the 28th International Conference on Machine
  Learning (ICML)}, pages 361--368, 2011.

\bibitem[Chung and Cho(2022)]{Chung_Cho_2022}
W.~Chung and Y.~Cho.
\newblock Bayesian mixed models for longitudinal genetic data: Theory,
  concepts, and simulation studies.
\newblock \emph{Genomics \& Informatics}, 20\penalty0 (1), 2022.

\bibitem[Daniels and Hogan(2008)]{Daniels_Hogan_2008}
Michael~J. Daniels and Joseph~W. Hogan.
\newblock \emph{Missing Data in Longitudinal Studies: Strategies for Bayesian
  Modeling and Sensitivity Analysis}.
\newblock Chapman \& Hall/CRC, 2008.

\bibitem[Diggle et~al.(2002)Diggle, Heagerty, Liang, and Zeger]{Diggle_2002}
Peter~J. Diggle, Patrick Heagerty, Kung-Yee Liang, and Scott~L. Zeger.
\newblock \emph{Analysis of Longitudinal Data}.
\newblock Oxford University Press, 2 edition, 2002.

\bibitem[Dunson(2010)]{Dunson_2010}
David~B. Dunson.
\newblock Nonparametric bayes applications to biostatistics.
\newblock In Nils~Lid Hjort, Chris Holmes, Peter M{\"u}ller, and Stephen~G.
  Walker, editors, \emph{Bayesian Nonparametrics}, Cambridge Series in
  Statistical and Probabilistic Mathematics, pages 223--273. Cambridge
  University Press, 2010.

\bibitem[Efron(2012)]{Efron_2012}
Bradley Efron.
\newblock \emph{Large-Scale Inference: Empirical Bayes Methods for Estimation,
  Testing, and Prediction}.
\newblock IMS Monographs. Cambridge University Press, 2012.

\bibitem[Efron and Morris(1971)]{Efron_Morris_1971}
Bradley Efron and Carl Morris.
\newblock Limiting the risk of bayes and empirical bayes estimators -- part i:
  The bayes case.
\newblock \emph{Journal of the American Statistical Association}, 66:\penalty0
  807--815, 1971.

\bibitem[Efron and Morris(1972)]{Efron_Morris_1972}
Bradley Efron and Carl Morris.
\newblock Limiting the risk of bayes and empirical bayes estimators -- part ii:
  The empirical bayes case.
\newblock \emph{Journal of the American Statistical Association}, 67:\penalty0
  130--139, 1972.

\bibitem[Egerton and Laycock(1982)]{Egerton_Laycock_1982}
M.~F. Egerton and P.~J. Laycock.
\newblock An explicit formula for the risk of james{-}{-}stein estimators.
\newblock \emph{Canadian Journal of Statistics}, 10:\penalty0 199--205, 1982.

\bibitem[Finn et~al.(2017)Finn, Abbeel, and Levine]{Finn_Abbeel_Levine_2017}
Chelsea Finn, Pieter Abbeel, and Sergey Levine.
\newblock Model-agnostic meta-learning for fast adaptation of deep networks.
\newblock In \emph{Proceedings of the 34th International Conference on Machine
  Learning (ICML)}, volume~70. PMLR, 2017.

\bibitem[Fitzmaurice et~al.(2011)Fitzmaurice, Laird, and
  Ware]{Fitzmaurice_Laird_Ware_2011}
Garrett~M. Fitzmaurice, Nan~M. Laird, and James~H. Ware.
\newblock \emph{Applied Longitudinal Analysis}.
\newblock Wiley, 2 edition, 2011.

\bibitem[Gelman and Hill(2007)]{Gelman_Hill_2007}
Andrew Gelman and Jennifer Hill.
\newblock \emph{Data Analysis Using Regression and Multilevel/Hierarchical
  Models}.
\newblock Cambridge University Press, 2007.

\bibitem[Gelman et~al.(2014)Gelman, Carlin, Stern, Dunson, Vehtari, and
  Rubin]{Gelman_etal_2014}
Andrew Gelman, John~B. Carlin, Hal~S. Stern, David~B. Dunson, Aki Vehtari, and
  Donald~B. Rubin.
\newblock \emph{Bayesian Data Analysis}.
\newblock Texts in Statistical Science. CRC Press, 3 edition, 2014.

\bibitem[Goel(1983)]{Goel_1983}
Prasanta~K. Goel.
\newblock Information measures and bayesian hierarchical models.
\newblock \emph{Journal of the American Statistical Association}, 78:\penalty0
  408--410, 1983.

\bibitem[Goel and DeGroot(1979)]{Goel_Degroot_1979}
Prasanta~K. Goel and Morris~H. DeGroot.
\newblock Comparison of experiments and information measures.
\newblock \emph{Annals of Statistics}, 7:\penalty0 1066--1077, 1979.

\bibitem[Goel and DeGroot(1981)]{Goel_Degroot_1981}
Prasanta~K. Goel and Morris~H. DeGroot.
\newblock Information about hyperparameters in hierarchical models.
\newblock \emph{Journal of the American Statistical Association}, 76:\penalty0
  140--147, 1981.

\bibitem[Goldstein(2011)]{Goldstein_2011}
Harvey Goldstein.
\newblock \emph{Multilevel Statistical Models}.
\newblock Wiley, Chichester, 4 edition, 2011.

\bibitem[Good(1950)]{Good_1950}
I.~J. Good.
\newblock \emph{Probability and Weighing of Evidence}.
\newblock Hafner, New York, 1950.

\bibitem[Good(1980)]{Good_1980}
I.~J. Good.
\newblock Some history of the hierarchical bayesian methodology.
\newblock In J.~Bernardo and M.~H. DeGroot, editors, \emph{Proceedings of the
  International Meeting on Bayesian Statistics}, Valencia, 1980. University of
  Valencia.

\bibitem[Good et~al.(1966)Good, Hacking, Jeffrey, and
  T{\"o}rnebohm]{Good_etal_1966}
I.~J. Good, I.~Hacking, R.~C. Jeffrey, and H.~T{\"o}rnebohm.
\newblock \emph{The Estimation of Probabilities: An Essay on Modern Bayesian
  Methods}.
\newblock MIT Press, Cambridge, Massachusetts, 1966.

\bibitem[Gordon et~al.(2019)Gordon, Bronskill, Bauer, Nowozin, and
  Turner]{Gordon_etal_2019}
Jonathan Gordon, John Bronskill, Matthias Bauer, Sebastian Nowozin, and
  Richard~E. Turner.
\newblock Meta-learning probabilistic inference for prediction.
\newblock In \emph{7th International Conference on Learning Representations
  (ICLR)}, New Orleans, LA, USA, 2019.

\bibitem[Grant et~al.(2018)Grant, Finn, Levine, Darrell, and
  Griffiths]{Grant_etal_2018}
Erin Grant, Chelsea Finn, Sergey Levine, Trevor Darrell, and Thomas~L.
  Griffiths.
\newblock Recasting gradient-based meta-learning as hierarchical bayes.
\newblock arXiv preprint, 2018.

\bibitem[Griffin and Holmes(2010)]{Griffin_Holmes_2010}
Jim Griffin and Chris Holmes.
\newblock Computational issues arising in bayesian nonparametric hierarchical
  models.
\newblock In Nils~Lid Hjort, Chris Holmes, Peter M{\"u}ller, and Stephen~G.
  Walker, editors, \emph{Bayesian Nonparametrics}, Cambridge Series in
  Statistical and Probabilistic Mathematics, pages 208--222. Cambridge
  University Press, 2010.

\bibitem[Hedeker and Gibbons(2006)]{Hedeker_Gibbons_2006}
Donald Hedeker and Robert~D. Gibbons.
\newblock \emph{Longitudinal Data Analysis}.
\newblock Wiley, Hoboken, NJ, 2006.

\bibitem[Hox et~al.(2017)Hox, Moerbeek, and Van~de Schoot]{Hox_etal_2017}
Joop~J. Hox, Mirjam Moerbeek, and Rens Van~de Schoot.
\newblock \emph{Multilevel Analysis: Techniques and Applications}.
\newblock Routledge, New York, 3 edition, 2017.

\bibitem[Hu et~al.(2019)Hu, Moreno, Xiao, Shen, Obozinski, Lawrence, and
  Damianou]{Hu_etal_2019}
Shengyang Hu, Pedro~G. Moreno, Yingzhen Xiao, Xiaoyu Shen, Guillaume Obozinski,
  Neil Lawrence, and Andreas Damianou.
\newblock Empirical bayes transductive meta-learning with synthetic gradients.
\newblock In \emph{7th International Conference on Learning Representations
  (ICLR)}, New Orleans, LA, USA, 2019.

\bibitem[James and Stein(1961)]{James_Stein_1961}
W.~James and C.~Stein.
\newblock Estimation with quadratic loss.
\newblock In \emph{Proceedings of the Fourth Berkeley Symposium on Mathematical
  Statistics and Probability}, volume~1, pages 361--380, Berkeley, 1961.
  University of California Press.

\bibitem[Johnstone and Silverman(2004)]{Johnstone_Silverman_2004}
Iain Johnstone and Bernard~W. Silverman.
\newblock Needles and straw in haystacks: Empirical-bayes estimates of possibly
  sparse sequences.
\newblock \emph{Annals of Statistics}, 32:\penalty0 1594--1649, 2004.

\bibitem[Karbalayghareh et~al.(2018)Karbalayghareh, Qian, and
  Dougherty]{Karbalayghareh_etal_2018}
Alireza Karbalayghareh, Xiaoning Qian, and Edward~R. Dougherty.
\newblock Optimal bayesian transfer learning.
\newblock \emph{IEEE Transactions on Signal Processing}, 66:\penalty0
  3724--3739, 2018.

\bibitem[Kong et~al.(2020)Kong, Somani, Song, Kakade, and Oh]{Kong_etal_2020}
Weiyang Kong, Raghav Somani, Zhao Song, Sham Kakade, and Sewoong Oh.
\newblock Meta-learning for mixed linear regression.
\newblock In \emph{Proceedings of the 37th International Conference on Machine
  Learning (ICML)}, volume 119. PMLR, 2020.

\bibitem[Laird and Ware(1982)]{Laird_Ware_1982}
N.~M. Laird and J.~H. Ware.
\newblock Random-effects models for longitudinal data.
\newblock \emph{Biometrics}, 38:\penalty0 963--974, 1982.

\bibitem[Lehmann and Casella(2006)]{Lehmann_Casella_2006}
E.~L. Lehmann and George Casella.
\newblock \emph{Theory of Point Estimation}.
\newblock Springer Texts in Statistics. Springer, New York, 2006.

\bibitem[Lindley and Smith(1972)]{Lindley_Smith_1972}
D.~V. Lindley and A.~F.~M. Smith.
\newblock Bayes estimates for the linear model.
\newblock \emph{Journal of the Royal Statistical Society: Series B},
  34:\penalty0 1--41, 1972.

\bibitem[Lindsey(1993)]{Lindsey_1993}
James~K. Lindsey.
\newblock \emph{Models for Repeated Measurements}.
\newblock Oxford University Press, Oxford, 1993.

\bibitem[Mallows(1972)]{Mallows_1972}
Colin Mallows.
\newblock A note on asymptotic joint normality.
\newblock \emph{Annals of Mathematical Statistics}, 43:\penalty0 508--515,
  1972.

\bibitem[Molenberghs and Verbeke(2005)]{Molenberghs_Verbeke_2005}
Geert Molenberghs and Geert Verbeke.
\newblock \emph{Models for Discrete Longitudinal Data}.
\newblock Springer, New York, 2005.

\bibitem[Molenberghs et~al.(2010)Molenberghs, Verbeke, Dem{\'e}trio, and
  Vieira]{Molenberghs_etal_2010}
Geert Molenberghs, Geert Verbeke, Clarice G.~B. Dem{\'e}trio, and A.~M. Vieira.
\newblock A family of generalized linear models for repeated measures with
  normal and conjugate random effects.
\newblock \emph{Statistical Science}, 25\penalty0 (3):\penalty0 325--347, 2010.

\bibitem[Nguyen(2016)]{Nguyen_2016}
XuanLong Nguyen.
\newblock Borrowing strength in hierarchical bayes: Posterior concentration of
  the dirichlet base measure.
\newblock \emph{Bernoulli}, 22:\penalty0 1535--1571, 2016.

\bibitem[Ohlssen et~al.(2007)Ohlssen, Sharples, and
  Spiegelhalter]{Ohlssen_2007}
David~I. Ohlssen, Linda~D. Sharples, and David~J. Spiegelhalter.
\newblock Flexible random-effects models using bayesian semi-parametric models:
  Applications to institutional comparisons.
\newblock \emph{Statistics in Medicine}, 26:\penalty0 2088--2112, 2007.

\bibitem[Raudenbush and Bryk(2002)]{Raudenbush_Bryk_2002}
Stephen~W. Raudenbush and Anthony~S. Bryk.
\newblock \emph{Hierarchical Linear Models: Applications and Data Analysis
  Methods}.
\newblock Sage, Thousand Oaks, 2 edition, 2002.

\bibitem[Ravi and Beatson(2019)]{Ravi_Beatson_2019}
Sachin Ravi and Alex Beatson.
\newblock Amortized bayesian meta-learning.
\newblock In \emph{7th International Conference on Learning Representations
  (ICLR)}, New Orleans, LA, USA, 2019.

\bibitem[Robert(2007)]{Robert_2007}
Christian~P. Robert.
\newblock \emph{The Bayesian Choice: From Decision-Theoretic Foundations to
  Computational Implementation}.
\newblock Springer Texts in Statistics. Springer, 2 edition, 2007.

\bibitem[Salakhutdinov et~al.(2013)Salakhutdinov, Tenenbaum, and
  Torralba]{Salakhutdinov_etal_2013}
Ruslan Salakhutdinov, Joshua~B. Tenenbaum, and Antonio Torralba.
\newblock A bayesian mixed-effects model to learn trajectories of changes from
  repeated manifold-valued observations.
\newblock \emph{IEEE Transactions on Pattern Analysis and Machine
  Intelligence}, 35, 2013.

\bibitem[Schiratti et~al.(2017)Schiratti, Allassonni{\`e}re, Colliot, and
  Durrleman]{Schiratti_etal_2017}
Jean-Baptiste Schiratti, St{\'e}phanie Allassonni{\`e}re, Olivier Colliot, and
  Stanley Durrleman.
\newblock Learning with hierarchical-deep models.
\newblock \emph{Journal of Machine Learning Research}, 18:\penalty0 1--33,
  2017.

\bibitem[Snijders and Bosker(2012)]{Snijders_Bosker_2012}
Tom A.~B. Snijders and Roel~J. Bosker.
\newblock \emph{Multilevel Analysis: An Introduction to Basic and Advanced
  Multilevel Modeling}.
\newblock Sage, London, 2 edition, 2012.

\bibitem[Stein(1981)]{Stein_1981}
Charles Stein.
\newblock Estimation of the mean of a multivariate normal distribution.
\newblock \emph{Annals of Statistics}, 9:\penalty0 1135--1151, 1981.

\bibitem[Teh et~al.(2005)Teh, Jordan, Beal, and Blei]{Teh_etal_2005}
Yee~Whye Teh, Michael~I. Jordan, Matthew~J. Beal, and David~M. Blei.
\newblock Sharing clusters among related groups: Hierarchical dirichlet
  processes.
\newblock In \emph{Advances in Neural Information Processing Systems}, pages
  1385--1392, 2005.

\bibitem[Vehtari et~al.(2017)Vehtari, Gelman, and Gabry]{Vehtari_etal_2017}
Aki Vehtari, Andrew Gelman, and Jonah Gabry.
\newblock Practical bayesian model evaluation using leave-one-out
  cross-validation and waic.
\newblock \emph{Statistics and Computing}, 27\penalty0 (5):\penalty0
  1413--1432, 2017.

\bibitem[Verbeke and Molenberghs(2009)]{Verbeke_Molenberghs_2009}
Geert Verbeke and Geert Molenberghs.
\newblock \emph{Linear Mixed Models for Longitudinal Data}.
\newblock Springer Science \& Business Media, 2009.

\bibitem[Verbeke et~al.(2014)Verbeke, Fieuws, Molenberghs, and
  Davidian]{Verbeke_etal_2014}
Geert Verbeke, Steffen Fieuws, Geert Molenberghs, and Marie Davidian.
\newblock The analysis of multivariate longitudinal data: A review.
\newblock \emph{Statistical Methods in Medical Research}, 23\penalty0
  (1):\penalty0 42--59, 2014.

\bibitem[Watanabe(2010)]{Watanabe_2010}
Sumio Watanabe.
\newblock Asymptotic equivalence of bayes cross validation and widely
  applicable information criterion in singular learning theory.
\newblock \emph{Journal of Machine Learning Research}, 11:\penalty0 3571--3594,
  2010.

\bibitem[Xu et~al.(2020)Xu, Zhu, and Lee]{Xu_etal_2020}
Guang Xu, Hongtu Zhu, and Jason~J. Lee.
\newblock Borrowing strength and borrowing index for bayesian hierarchical
  models.
\newblock \emph{Computational Statistics and Data Analysis}, 144, 2020.

\bibitem[Yoon et~al.(2018)Yoon, Kim, Dia, Kim, Bengio, and Ahn]{Yoon_etal_2018}
Jaehoon Yoon, Taesup Kim, Ousmane Dia, Sungjin Kim, Yoshua Bengio, and Sungjin
  Ahn.
\newblock Bayesian model-agnostic meta-learning.
\newblock In \emph{Advances in Neural Information Processing Systems
  (NeurIPS)}, Montr{\'e}al, Canada, 2018.

\bibitem[Zdeborov{\'a} and Krzakala(2016)]{Zdeborova_2016}
Lenka Zdeborov{\'a} and Florent Krzakala.
\newblock Statistical physics of inference: Thresholds and algorithms.
\newblock \emph{Advances in Physics}, 65\penalty0 (5):\penalty0 453--552, 2016.

\end{thebibliography}


\end{document}